\documentclass[prc,twocolumn,superscriptaddress,showpacs,amssymb,amsmath,amsfonts,aps]{revtex4}
\usepackage{graphicx}
\usepackage{dcolumn}
\usepackage{epsfig}
\begin{document}
\title{Measurements of ${\rm\bf {ep \to {\rm\bf e^\prime \pi^+ n}}}$ at 
${\rm\bf 1.6 < W < 2.0}$~GeV and extraction of nucleon resonance electrocouplings at CLAS}

\newcommand*{\ASU}{Arizona State University, Tempe, Arizona 85287-1504, USA}
\newcommand*{\ASUindex}{1}
\affiliation{\ASU}
\newcommand*{\CSUDH}{California State University, Dominguez Hills, Carson, California 90747, USA}
\newcommand*{\CSUDHindex}{2}
\affiliation{\CSUDH}
\newcommand*{\CANISIUS}{Canisius College, Buffalo, New York, USA}
\newcommand*{\CANISIUSindex}{3}
\affiliation{\CANISIUS}
\newcommand*{\CMU}{Carnegie Mellon University, Pittsburgh, Pennsylvania 15213, USA}
\newcommand*{\CMUindex}{4}
\affiliation{\CMU}
\newcommand*{\CUA}{Catholic University of America, Washington, D.C. 20064, USA}
\newcommand*{\CUAindex}{5}
\affiliation{\CUA}
\newcommand*{\SACLAY}{CEA, Centre de Saclay, Irfu/Service de Physique Nucl\'eaire, 91191 Gif-sur-Yvette, France}
\newcommand*{\SACLAYindex}{6}
\affiliation{\SACLAY}
\newcommand*{\CNU}{Christopher Newport University, Newport News, Virginia 23606, USA}
\newcommand*{\CNUindex}{7}
\affiliation{\CNU}
\newcommand*{\UCONN}{University of Connecticut, Storrs, Connecticut 06269, USA}
\newcommand*{\UCONNindex}{8}
\affiliation{\UCONN}
\newcommand*{\FU}{Fairfield University, Fairfield Connecticut 06824, USA}
\newcommand*{\FUindex}{8}
\affiliation{\FU}
\newcommand*{\FIU}{Florida International University, Miami, Florida 33199, USA}
\newcommand*{\FIUindex}{9}
\affiliation{\FIU}
\newcommand*{\FSU}{Florida State University, Tallahassee, Florida 32306, USA}
\newcommand*{\FSUindex}{10}
\affiliation{\FSU}
\newcommand*{\Genova}{Universit$\grave{a}$ di Genova, 16146 Genova, Italy}
\newcommand*{\Genovaindex}{11}
\affiliation{\Genova}
\newcommand*{\GWUI}{The George Washington University, Washington, DC 20052, USA}
\newcommand*{\GWUIindex}{12}
\affiliation{\GWUI}
\newcommand*{\ISU}{Idaho State University, Pocatello, Idaho 83209, USA}
\newcommand*{\ISUindex}{13}
\affiliation{\ISU}
\newcommand*{\INFNFE}{INFN, Sezione di Ferrara, 44100 Ferrara, Italy}
\newcommand*{\INFNFEindex}{14}
\affiliation{\INFNFE}
\newcommand*{\INFNFR}{INFN, Laboratori Nazionali di Frascati, 00044 Frascati, Italy}
\newcommand*{\INFNFRindex}{15}
\affiliation{\INFNFR}
\newcommand*{\INFNGE}{INFN, Sezione di Genova, 16146 Genova, Italy}
\newcommand*{\INFNGEindex}{16}
\affiliation{\INFNGE}
\newcommand*{\INFNRO}{INFN, Sezione di Roma Tor Vergata, 00133 Rome, Italy}
\newcommand*{\INFNROindex}{17}
\affiliation{\INFNRO}
\newcommand*{\INFNTUR}{INFN, sez. di Torino, 10125 Torino, Italy}
\newcommand*{\INFNTURindex}{18}
\affiliation{\INFNTUR}
\newcommand*{\ORSAY}{Institut de Physique Nucl\'eaire ORSAY, Orsay, France}
\newcommand*{\ORSAYindex}{19}
\affiliation{\ORSAY}
\newcommand*{\ITEP}{Institute of Theoretical and Experimental Physics, Moscow, 117259, Russia}
\newcommand*{\ITEPindex}{20}
\affiliation{\ITEP}
\newcommand*{\JMU}{James Madison University, Harrisonburg, Virginia 22807, USA}
\newcommand*{\JMUindex}{21}
\affiliation{\JMU}
\newcommand*{\KNU}{Kyungpook National University, Daegu 702-701, Republic of Korea}
\newcommand*{\KNUindex}{22}
\affiliation{\KNU}
\newcommand*{\LPSC}{LPSC, Universit\'e Grenoble-Alpes, CNRS/IN2P3, Grenoble, France}
\newcommand*{\LPSCindex}{23}
\affiliation{\LPSC}
\newcommand*{\UNH}{University of New Hampshire, Durham, New Hampshire 03824-3568, USA}
\newcommand*{\UNHindex}{24}
\affiliation{\UNH}
\newcommand*{\NSU}{Norfolk State University, Norfolk, Virginia 23504, USA}
\newcommand*{\NSUindex}{25}
\affiliation{\NSU}
\newcommand*{\OHIOU}{Ohio University, Athens, Ohio  45701, USA}
\newcommand*{\OHIOUindex}{26}
\affiliation{\OHIOU}
\newcommand*{\ODU}{Old Dominion University, Norfolk, Virginia 23529, USA}
\newcommand*{\ODUindex}{27}
\affiliation{\ODU}
\newcommand*{\RPI}{Rensselaer Polytechnic Institute, Troy, New York 12180-3590, USA}
\newcommand*{\RPIindex}{28}
\affiliation{\RPI}
\newcommand*{\URICH}{University of Richmond, Richmond, Virginia 23173, USA}
\newcommand*{\URICHindex}{29}
\affiliation{\URICH}
\newcommand*{\ROMAII}{Universita' di Roma Tor Vergata, 00133 Rome Italy}
\newcommand*{\ROMAIIindex}{31}
\affiliation{\ROMAII}
\newcommand*{\MSU}{Skobeltsyn Institute of Nuclear Physics, Lomonosov Moscow State University, 119234 Moscow, Russia}
\newcommand*{\MSUindex}{30}
\affiliation{\MSU}
\newcommand*{\SCAROLINA}{University of South Carolina, Columbia, South Carolina 29208, USA}
\newcommand*{\SCAROLINAindex}{31}
\affiliation{\SCAROLINA}
\newcommand*{\TEMPLE}{Temple University,  Philadelphia, PA 19122, USA }
\newcommand*{\TEMPLEindex}{32}
\affiliation{\TEMPLE}
\newcommand*{\JLAB}{Thomas Jefferson National Accelerator Facility, Newport News, Virginia 23606, USA}
\newcommand*{\JLABindex}{33}
\affiliation{\JLAB}
\newcommand*{\UTFSM}{Universidad T\'{e}cnica Federico Santa Mar\'{i}a, Casilla 110-V Valpara\'{i}so, Chile}
\newcommand*{\UTFSMindex}{34}
\affiliation{\UTFSM}
\newcommand*{\EDINBURGH}{Edinburgh University, Edinburgh EH9 3JZ, United Kingdom}
\newcommand*{\EDINBURGHindex}{35}
\affiliation{\EDINBURGH}
\newcommand*{\GLASGOW}{University of Glasgow, Glasgow G12 8QQ, United Kingdom}
\newcommand*{\GLASGOWindex}{36}
\affiliation{\GLASGOW}
\newcommand*{\VIRGINIA}{University of Virginia, Charlottesville, Virginia 22901, USA}
\newcommand*{\VIRGINIAindex}{37}
\affiliation{\VIRGINIA}
\newcommand*{\WM}{College of William and Mary, Williamsburg, Virginia 23187-8795, USA}
\newcommand*{\WMindex}{38}
\affiliation{\WM}
\newcommand*{\YEREVAN}{Yerevan Physics Institute, 375036 Yerevan, Armenia}
\newcommand*{\YEREVANindex}{39}
\affiliation{\YEREVAN}
\newcommand*{\ANL}{Argonne National Laboratory, Argonne, Illinois 60439, USA}
\newcommand*{\ANLindex}{40}
\affiliation{\ANL}

\newcommand*{\VT}{Virginia Polytechnic Institute and State University, Blacksburg, Virginia 24061-0435, USA}
\newcommand*{\VTindex}{41}
\affiliation{\VT}
\newcommand*{\MSUM}{M.V.Lomonosov Moscow State University,Leninskie Gory, Moscow 119991, Russia}
\newcommand*{\MSUMindex}{42}
\affiliation{\MSUM}

\newcommand*{\NOWCNU}{Christopher Newport University, Newport News, Virginia 23606, USA}
\newcommand*{\NOWJLAB}{Thomas Jefferson National Accelerator Facility, Newport News, Virginia 23606, USA}
\newcommand*{\NOWUCONN}{University of Connecticut, Storrs, Connecticut 06269, USA}
\newcommand*{\NOWGLASGOW}{University of Glasgow, Glasgow G12 8QQ, United Kingdom}
\newcommand*{\NOWINFNGE}{INFN, Sezione di Genova, 16146 Genova, Italy}
\newcommand*{\NOWMSSU}{Mississippi State University, 125 Hilbun Hall, Miss State, Mississippi 39762, USA}
 %%%%%%%%%%%%%%% END OF Latex Macros for institute addresses  %%%%%%%%%%%%%%%%%%%%%%%%% 

\author {K.~Park} 
\affiliation{\JLAB}
\affiliation{\ODU}
\author {I.G.~Aznauryan} 
\affiliation{\JLAB}
\affiliation{\YEREVAN}
\author {V.D.~Burkert} 
\affiliation{\JLAB}

\author {K.P. ~Adhikari} 
\affiliation{\ODU}
\author {M.J.~Amaryan} 
\affiliation{\ODU}
\author {S. ~Anefalos~Pereira} 
\affiliation{\INFNFR}
\author {H. ~Avakian} 
\affiliation{\JLAB}
\author {M.~Battaglieri} 
\affiliation{\INFNGE}
\author {R.~Badui} 
\affiliation{\FIU}
\author {I.~Bedlinskiy} 
\affiliation{\ITEP}
% ...

\collaboration{The CLAS Collaboration}
\noaffiliation

\vskip 1.cm

\begin{abstract}

{Differential cross sections of the exclusive process $e p \to e^\prime \pi^+ n$ were measured  
with good precision in the range of the photon virtuality $Q^2 = 1.8 - 4.5$~GeV$^2$, and the invariant 
mass range of the $\pi^+ n$ final state $W = 1.6 - 2.0$~GeV using the CEBAF Large 
Acceptance Spectrometer. Data were collected with nearly complete coverage in  
 the azimuthal and polar angles of the $n\pi^+$ center-of-mass system. 
More than 37,000 cross section points were 
measured. The contributions of the isospin $I = {1\over 2}$ 
resonances $N(1675){5\over 2}^-$,  $N(1680){5\over 2}^+$ and $N(1710){1\over 2}^+$ 
were extracted at different values of $Q^2$ using a single-channel, energy-dependent
resonance amplitude analysis. Two different approaches, the unitary 
isobar model and the fixed-$t$ dispersion relations, were employed in
the analysis. We observe significant 
  strength of the $N(1675){5\over 2}^-$ in the $A_{1/2}$ amplitude, which is 
  in strong disagreement with quark models that predict both transverse amplitudes
  to be strongly suppressed. 
  For the $N(1680){5\over 2}^+$ we observe a slow changeover from the 
dominance of the $A_{3/2}$ amplitude at the real photon point ($Q^2=0$) 
  to a $Q^2$ where $A_{1/2}$ begins to dominate. 
  The scalar amplitude $S_{1/2}$ drops rapidly with $Q^2$ consistent with quark 
  model prediction.  For the $N(1710){1\over2}^+$ resonance our analysis 
  shows significant strength for the $A_{1/2}$ amplitude at $Q^2 < 2.5$~GeV$^2$.}

\end{abstract}

\maketitle

\section{Introduction}

The study of the excited states of the nucleon is an important step in the development 
of a fundamental understanding of the strong interaction~\cite{isgur_2000}. While the 
existing data on the low-lying resonances are consistent with the well-studied 
$SU(6) \otimes O(3)$ constituent quark model classification, many open questions 
remain. On a fundamental level there exists only a very limited understanding of 
the relationship between Quantum Chromo-Dynamics (QCD), the field theory of the 
strong interaction, and the constituent quark model (CQM) or alternative hadron
 models, however recent developments in Lattice QCD, most notably the predictions 
 of the spectrum of $N^*$ and $\Delta^*$ states,
 have shown~\cite{Edwards:2011jj} that the same symmetry of $SU(6) \otimes O(3)$ is 
 likely at work here as is underlying the spectrum in the CQM.  
 
Experimentally, we still do not have sufficiently complete data that can be used to 
uncover unambiguously the structure of the nucleon and its excited states in the  
entire resonance mass range. While this remains an important long term goal, 
very significant advances have been made
during the past decade that have enabled the precise determination of resonance electrocouplings 
for a set of lower mass states and in a wide space-time range. 
Precise data have become available in recent years~\cite{Joo2002,Joo2003,Sparveris2003,Biselli2003,Kelly2005,Stave06,Ungaro06,Sparveris06}
to study the transition from the nucleon ground state to the $\Delta(1232)$, 
in $\pi^0$ electroproduction on the proton with wide angular coverage and in a wide 
range of four-momentum transfer $Q^2$. 
This has allowed for the determination of the magnetic dipole transition form factor and
the electric and scalar quadrupole transition, covering a range of 
$0 \leq Q^2 \leq 7~\rm{GeV^2}$ (we set $c=1$). 

This information,  combined with precise cross section and polarization data
 for the processes $ep \to e^\prime \pi^0 p$~\cite{Joo2002,Joo2003,Biselli2008}, 
$ep \to e^\prime \pi^+ n$~\cite{Joo2004,Egiyan:2006ks,Park:2007tn} and 
 $ep \to e^\prime \eta p$~\cite{Armstrong1999,Thompson2001,Denizli:2007tq} 
 in the second nucleon resonance region near $W = 1.35 - 1.6$~GeV allowed for precise 
 measurements of electrocouplings of the "Roper" resonance $N(1440){1\over 2}^+$~\cite{Aznauryan:2008pe}, which in the 
 CQM is the first radial excitation of the nucleon. These results solved a longstanding question 
 regarding the nature of this state.  Precise results have also been obtained 
 for the transition to the $N(1535){1\over 2}^-$ and the $N(1520){3\over 2}^-$ states.
 Following these breakthroughs, the process $ep\to ep\pi^+\pi^-$ was 
 measured in the lower $Q^2$ and low mass range~\cite{Fedotov:2008aa}, and a reaction 
 model was developed~\cite{Mokeev:2008iw} that enabled extraction of the electrocoupling 
 amplitudes for the resonances $N(1440){1\over 2}^+$ and $N(1520){3\over 2}^-$~\cite{Mokeev:2012vsa}
 from this channel. 
The two-pion results were consistent with the results from the single pion analysis, and thus validated 
the analysis approach for this more complex reaction channel. This is a highly non-trivial
result as the non-resonant (background) contributions are of completely different origin for the 
two processes.  

\begin{figure}[t]
\begin{center}
\epsfig{file=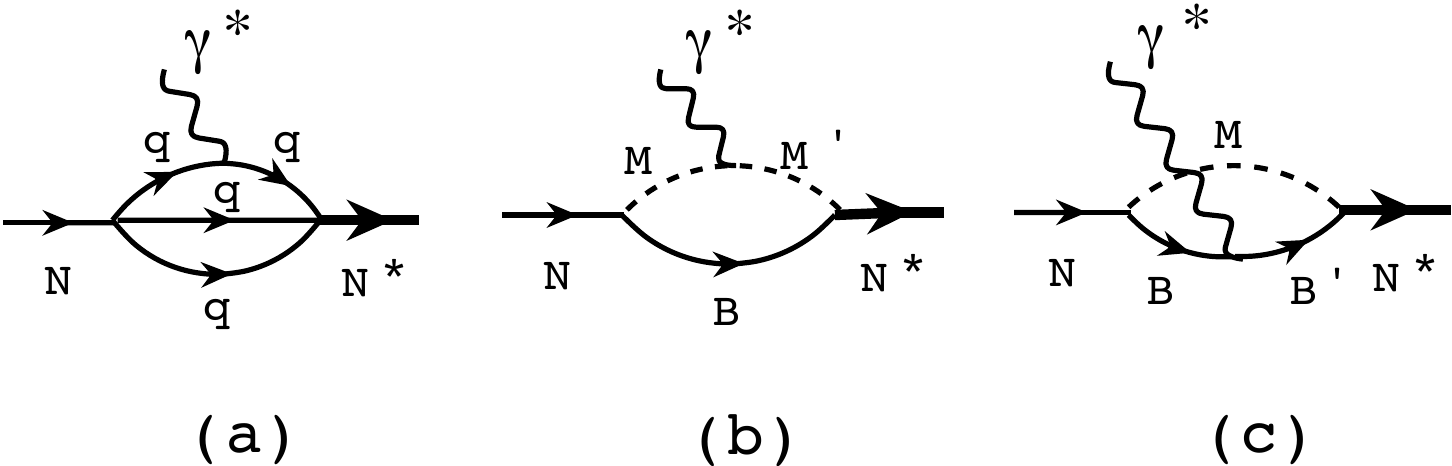,scale=0.58}
\caption{\small
Main contributions to the  $\gamma^* N \rightarrow N^*$
 transition: (a) through quark transition; (b,c) through meson-baryon pairs.
\label{meson}}
\end{center}
\end{figure}

The transition amplitudes for the lower mass excited states have been discussed extensively in recent 
reviews~\cite{Aznauryan2012,Aznauryan2013}. The progress is quite impressive 
when compared with results available before 2004~\cite{burkert_lee_2004}. 
 One of the major results of these analyses is the evidence for the need to include 
 significant meson-baryon contributions in models that describe the $Q^2$ dependence 
 of the resonance excitation strength. At low $Q^2$ these contributions 
 can be of the same magnitude as the quark contribution, but appear to fall off more rapidly with 
 increasing $Q^2$~\cite{Aznauryan:2012ec,Ramalho:2013mxa}. 
This information has been obtained largely through the observation that the 
quark transition processes  often do not have sufficient strength 
to explain fully the measured transition amplitudes. One of the best known 
examples  is the photoexcitation of the $\Delta(1232){3\over 2}^+$ on the proton.
This  reaction proceeds mostly through a magnetic dipole transition from the nucleon, 
but only about 70\% of the transition amplitude is explained by the quark content of the state.
A satisfactory description of the 
$\gamma^* p\rightarrow \Delta(1232){3\over 2}^+$ transition was achieved in
models that include pion-cloud contributions
\cite{Thomas,Faessler} and also in dynamical reaction models~\cite{Kamalov1999,Kamalov2001,Sato2001,Sato2007,Sato2008}, where
the missing strength has been attributed to dynamical meson-baryon interactions 
in the final state.
 Similar conclusions have been drawn for the excited nucleon states 
$N(1440){1\over 2}^+$, $N(1520){3\over 2}^-$ and 
$N(1535){1\over 2}^-$ using a constituent quark model on the light-cone~\cite{Aznauryan:2012ec},
and a relativistic quark model with spectator 
di-quark~\cite{Ramalho:2010js,Ramalho:2013mxa,Ramalho:2011ae}. 
The two main processes that contribute to the  $\gamma^* N \rightarrow N^*$ 
transition are illustrated in Fig.~\ref{meson}, by the diagrams (a) and (b,c). 

The focus of the current work is the study of the higher mass range $W >1.6$~GeV. 
Many $N^*$ and $\Delta^*$ resonances are known to populate this mass 
range~\cite{Agashe:2014kda}, and 
several of them couple strongly to the $N\pi$ final state and can be investigated with
the current study, while others couple more strongly to $N\pi\pi$ final states. 
In addition to the study of individual channels, a full exploration will require to   
analyze these channels together in a coupled-channel framework.
In this work we provide differential cross sections for the process 
$ep \to e^\prime  \pi^+ n$ in the range $1.6 <  W < 2.0$~GeV with nearly 
full azimuthal and polar angle coverage in the $\pi^+n$ system. In addition to providing
essential input to full coupled-channel analyses, we expect for some resonances, 
especially $N(1675){5\over 2}^-$ and $N(1680){5\over 2}^+$, that a 
single-channel analysis will yield reliable results due to the large coupling of these states 
to $N\pi$ and the absence of $I = {3\over 2}$ states with the same spin-parity in that 
mass range.

\section{Formalism}
\begin{figure}[thb]
\begin{center}
	\includegraphics[angle=0,width=0.45\textwidth]{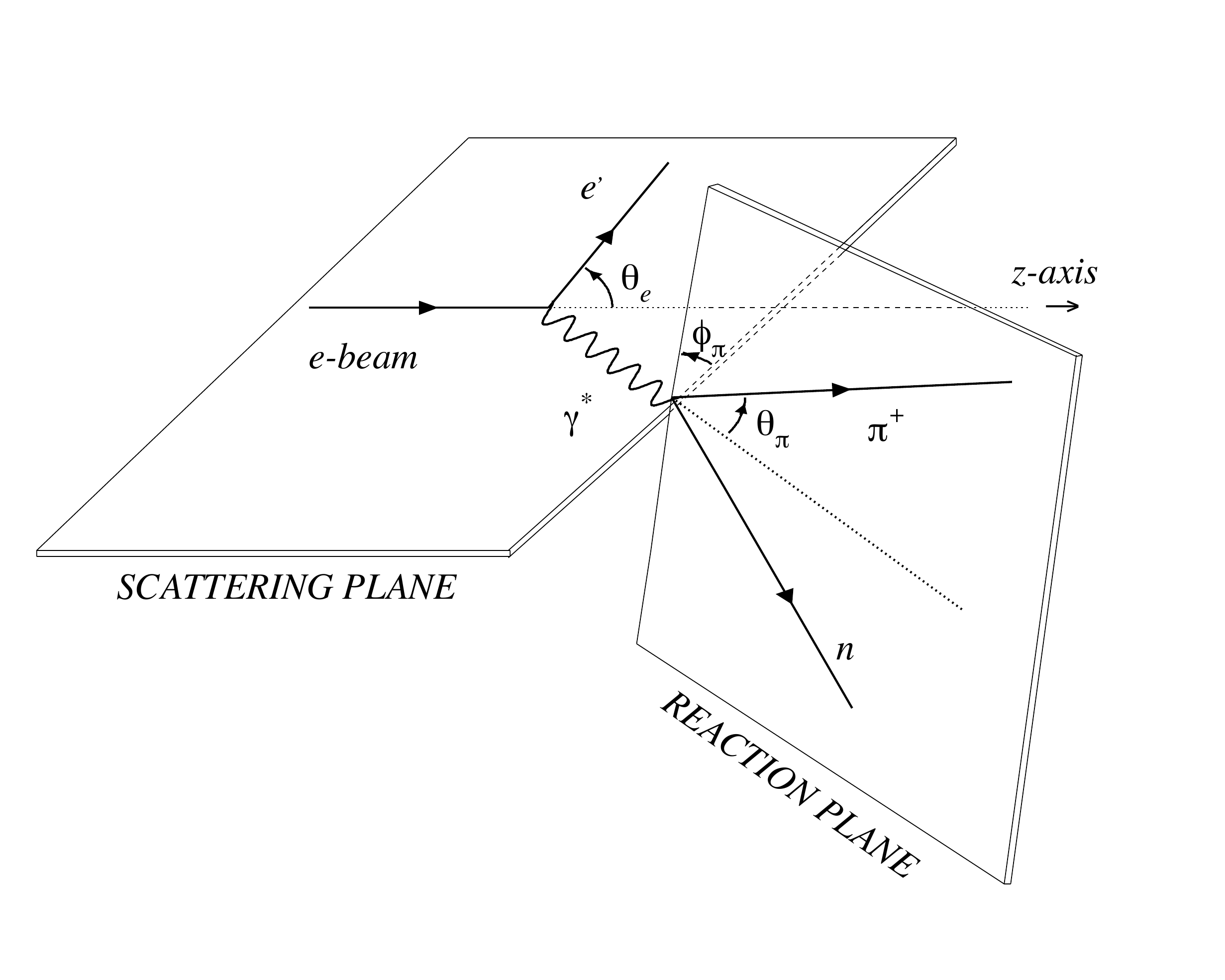}
        \caption{
          Kinematics of single $\pi^+$ electroproduction. }
          \label{fig:pion_kinematics}
\end{center}
\end{figure}

We report on measurements of differential cross sections with the CEBAF Large Acceptance Spectrometer (CLAS) at Jefferson Lab using a polarized continuous wave (CW) electron beam of 5.499 $\rm{GeV}$ energy incident upon a liquid-hydrogen target. The kinematics of single pion electroproduction is displayed in Fig.~\ref{fig:pion_kinematics}.
%%%%%%%%%%%%%%%%%%%%%%%%%%%%%%%%%%%%%%%%%%%%%%%%%%%%%%%%%%%%%%%%%
%%%%%%%%%%%%%%%%%%%%%%%%%%%%%%%%%%%%%%%%%%%%%%%%%%%%%%%%%%%%%%%%%
 In the one-photon exchange approximation the electron kinematics is described by two Lorentz invariants: 
$Q^2$, characterizing the virtuality of the exchanged photon, and $\nu$, the transferred energy: 
\begin{eqnarray} 
Q^2 \equiv -(k_i-k_f)^2 = 4E_iE_f\sin^2{\theta_e \over 2}~,\\
\nu \equiv \frac{p_i\cdot p_{\gamma}}{M_p} = E_i - E_f ~,
\end{eqnarray}
where $k_i$ and $k_f$ are the initial and final four-momenta of the electron, $p_{\gamma}$ and $p_i$ are the virtual photon and target four-momenta. $E_i$ and $E_f$ are the initial and final electron energies in the laboratory frame, $\theta_e$ is the electron scattering angle, and $M_p$ is the proton mass. Another related quantity is the invariant mass of the hadronic final state $W$ that can be expressed as: 
\begin{eqnarray}
W^2 \equiv (p_{\gamma} + p_i)^2 = M_p^2 + 2M_p\nu - Q^2~.
\end{eqnarray}        
In this measurement the scattered electron and the outgoing $\pi^+$ are detected while the final state neutron is unobserved. Since the four-momentum of the incident electron and of the target proton are known, the four-momentum of the missing system $X$ in the final state can be reconstructed and its mass determined as:  
\begin{eqnarray}
M_X^2 \equiv [(k_i + p_i) - (k_f + q_{\pi})]^2~,
\label{eqn:miss_mass}
\end{eqnarray}
where $q_{\pi}$ is the 4-momentum of the outgoing $\pi^+$. For single $\pi^+$ production, the constraint on the missing mass is $M_X = M_n$. The outgoing $\pi^+$ is defined by two angles in the center-of-mass (CM) frame, the polar angle $\theta^*_{\pi}$ and the azimuthal angle $\phi^*_{\pi}$. The latter is the angle between the electron scattering plane and the hadronic production plane. It is defined such that the scattered electron lies in the $\phi^*_{\pi} = 0$ half plane with the $z$-axis pointing along the virtual photon three-momentum vector.  The kinematics is completely defined by five variables $(Q^2, W, \theta^*_{\pi}, \phi^*_{\pi}, \phi_e)$, where $\phi_e$ is the electron azimuthal lab angle. In the absence of a transverse polarization of the beam or the target nucleon, the cross section does not depend on $\phi_e$, and can be written as~\cite{burkert_lee_2004}:
\begin{eqnarray}
\frac{d^5\sigma}{dE_f d\Omega_e d\Omega^*_{\pi}} &=& \Gamma \cdot \frac{d^2\sigma}{d\Omega^*_{\pi}}~,
\end{eqnarray}
where
\begin{eqnarray} 
\Gamma& =& \frac{\alpha}{2\pi^2Q^2} \frac{(W^2-M_p^2)E_f}{2M_pE_e} \frac{1}{1-\epsilon}~,\\
\epsilon &=& (1 + 2(1 + \frac{\nu^2}{Q^2})\tan^2{\theta_e\over 2})^{-1}~,     \\ \nonumber
\frac{d\sigma} {d\Omega_{\pi}^*}&=&\sigma_T+\epsilon \sigma_L+\epsilon \sigma_{TT}\cos{2\phi^*_{\pi}} + \sqrt{2\epsilon(1+\epsilon)} \sigma_{LT}\cos{\phi^*_{\pi}}~. \\
\label{eq:diffcrs}
\end{eqnarray}

The parameter $\epsilon$ represents the virtual photon polarization, $\Gamma$ is the virtual photon flux, and $\frac{d^2\sigma}{d\Omega^*_{\pi}}$ is the differential photoabsorption cross section.

\section{Experimental Setup} 
\begin{figure}[!htb]
\begin{center}
	\includegraphics[angle=0,width=0.48\textwidth]{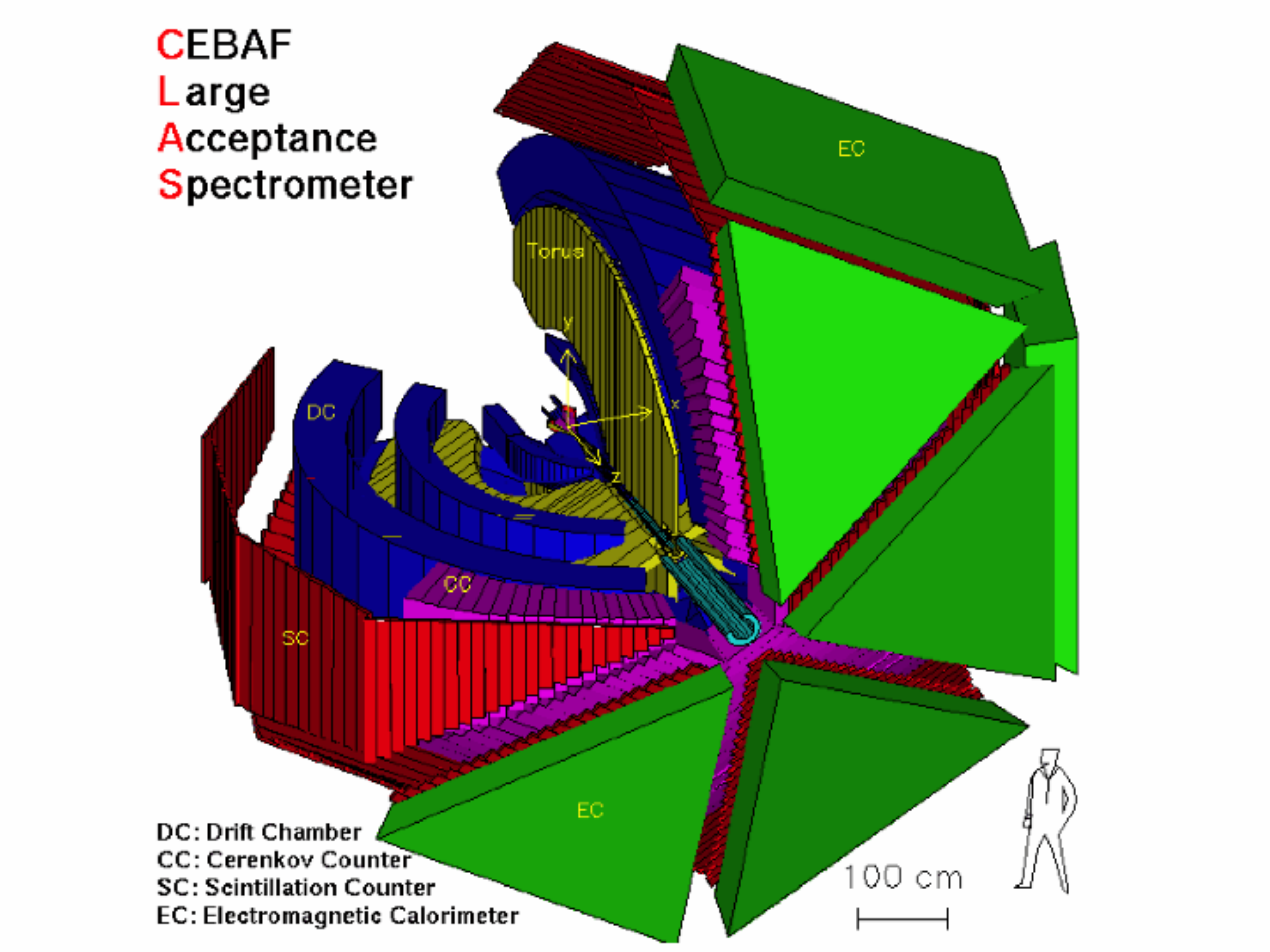}
        \caption{ (Color online) 
	 Cut view of the CLAS detector system. The beam enters from the upper left side into CLAS. 
	 The 6 superconducting torus magnet coils 
	 separate the detector into 6 independent spectrometers (sectors) each equipped with 3 regions 
	 of drift chambers. Time-of-flight scintillators cover the entire sector from polar angles of about 
	 $8^\circ$ to 140$^\circ$ and provide fast timing information for charged particle identification. 
	 In the forward angle range at polar
	  angles up to $\theta = 45^\circ$, the combination of gas $\check{\rm C}$erenkov counters and 
	  electromagnetic calorimeters provide electron identification and level 1 trigger capabilities. 
          \label{fig:clas}
        }
\end{center}
\end{figure}

The measurement was carried out using the CEBAF Large Acceptance Spectrometer (CLAS). Details of the detector systems and the operational performance of CLAS  are described elsewhere~\cite{clas}. A schematic view of CLAS is shown in Fig.~\ref{fig:clas}. CLAS utilizes a magnetic field distribution generated by six flat superconducting coils, arranged symmetrically in azimuth. The coils generate an approximate toroidal field distribution around the beam axis. The six identical sectors of the magnet are independently instrumented with 34 layers of drift cells for particle tracking, plastic scintillation counters for time-of-flight (TOF) measurements and charged particle identification, gas threshold $\check{\rm C}$erenkov counters (CC) for electron and pion separation, and scintillator-lead sampling calorimeters (EC) for photon and neutron detection. To aid in electron/pion separation, the EC is segmented into an inner part of about 
6 radiation lengths facing the target, and an outer part of 9 radiation lengths away from the target. 
The energy accumulated in these two parts are called EC$_{\rm inner}$ 
and EC$_{\rm outer}$, respectively. CLAS covers on average 80\% of the full 4$\pi$ solid angle for the detection of charged particles. Azimuthal angle acceptance is maximum at large polar angles and decreases at forward angles. Polar angle coverage ranges from about 8$^{\circ}$ to 140$^{\circ}$ for the detection of $\pi^+$. Electrons are detected in the CC and EC covering polar angles from approximately 20$^{\circ}$ to 55$^{\circ}$, this range being somewhat dependent on the momentum of the scattered electron. The target was located 25~cm upstream of the nominal CLAS center, surrounded by a small toroidal magnet with normal conducting coils that was used to shield the drift chambers closest to the target from the intense low energy electron background resulting from M{\"o}ller scattering processes in the target. 
%%%%%%%%%%%%%%%%%%%%%%%%%%%%%%%%%%%%%%%%%%%%%%%%%%%%%%%%%%%%%%%%%
%%%%%%%%%%%%%%%%%%%%%%%%%%%%%%%%%%%%%%%%%%%%%%%%%%%%%%%%%%%%%%%%
In the current experiment, only two charged particles need to be detected, the scattered electron and the produced $\pi^+$, while the full final state is reconstructed using four-momentum conservation constraints. The CW beam provided by CEBAF is well suited for measurements involving two or more final state particles in coincidence, leading to very small accidental coincidence contributions of  $< 10^{-3}$ for the instantaneous luminosity of $10^{34}$cm$^{-2}$s$^{-1}$ used in this measurement.   

The measurement was performed from April to July 2003 as part of the CLAS run period e1f. An electron beam of 7- 8~nA current and an energy of 5.499 $\rm{GeV}$ was directed onto a 5~cm long liquid-hydrogen target. The beam charge was integrated in a totally absorbing Faraday cup (FC). Empty target runs were performed to measure contributions from the target cell windows. An integrated luminosity of $L=20$~fb$^{-1}$ was accumulated, and a total of $4.3\times 10^9$ triggers were collected containing $0.65\times 10^9$ events with at least one scattered electron. To optimize the overall acceptance and resolution, the torus magnet current was set at 2250~A corresponding to 2/3 of its normal operating field strength. Events were triggered on a single electron candidate defined as a coincidence of the total energy deposited in one sector of the EC and a signal in the CC of the same sector. A minimum energy of 640 $\rm{MeV}$ deposited in one EC sector was required in the trigger. All events were first written to a RAID
  disk array, and later transferred to the tape silo of the Jefferson Lab computer center. Raw data were subjected to the calibration and reconstruction procedures that are part of the standard CLAS data analysis chain. The reaction studied in this work contributed only a fraction to the total event sample, and a more stringent event selection was applied to select events with one electron candidate and only one positively charged track. These events were subject to further selection criteria described in the following sections. 
\section{Data Analysis}
\subsection{Event selection}
\subsubsection{Electron identification}
Selection of electron candidates in CLAS at the level 1 trigger is achieved by requiring energy deposited in the EC and a CC hit in the same sector. Such an open trigger does not provide a stringent electron selection at the relatively high beam energy, and additional selection criteria must be applied in the offline event analysis. First, we require that the EC and CC hits are geometrically matched with a negatively charged track in the drift chambers (DC). Secondly, we employ the direct correlation between the energy deposited in the scintillator part of the calorimeter ($E_{\rm dep}$) and the momentum obtained in the track reconstruction in the magnetic field. About 30\% of the total energy deposited in the EC ($E_{\rm tot}$) is directly measured in the active scintillator material. This detectable portion of the EM shower is referred to as the sampling fraction ($\alpha$). The remaining 70\% of the energy is deposited mostly in the lead sheets that are interleaved between the scintillator sheets as showering material. A GEANT3
 ~\cite{geant3} based Monte Carlo simulation package (GSIM) was used to determine the EC response as a function of electron energy.  The sampling fraction is nearly energy-independent and for this experiment $\alpha \equiv E_{\rm dep}/E_{\rm tot} = 0.28$.  Lower values of $\alpha$ are observed in cases where electrons hit the calorimeter near the edges, and a fraction of the shower energy leaks out of the calorimeter volume. Such edge effects are eliminated by defining fiducial regions that assure full energy response as long as the electrons hit the calorimeter inside the fiducial regions.  

\begin{figure}[!thb]
\begin{center}
  \includegraphics[angle=0,width=0.46\textwidth]{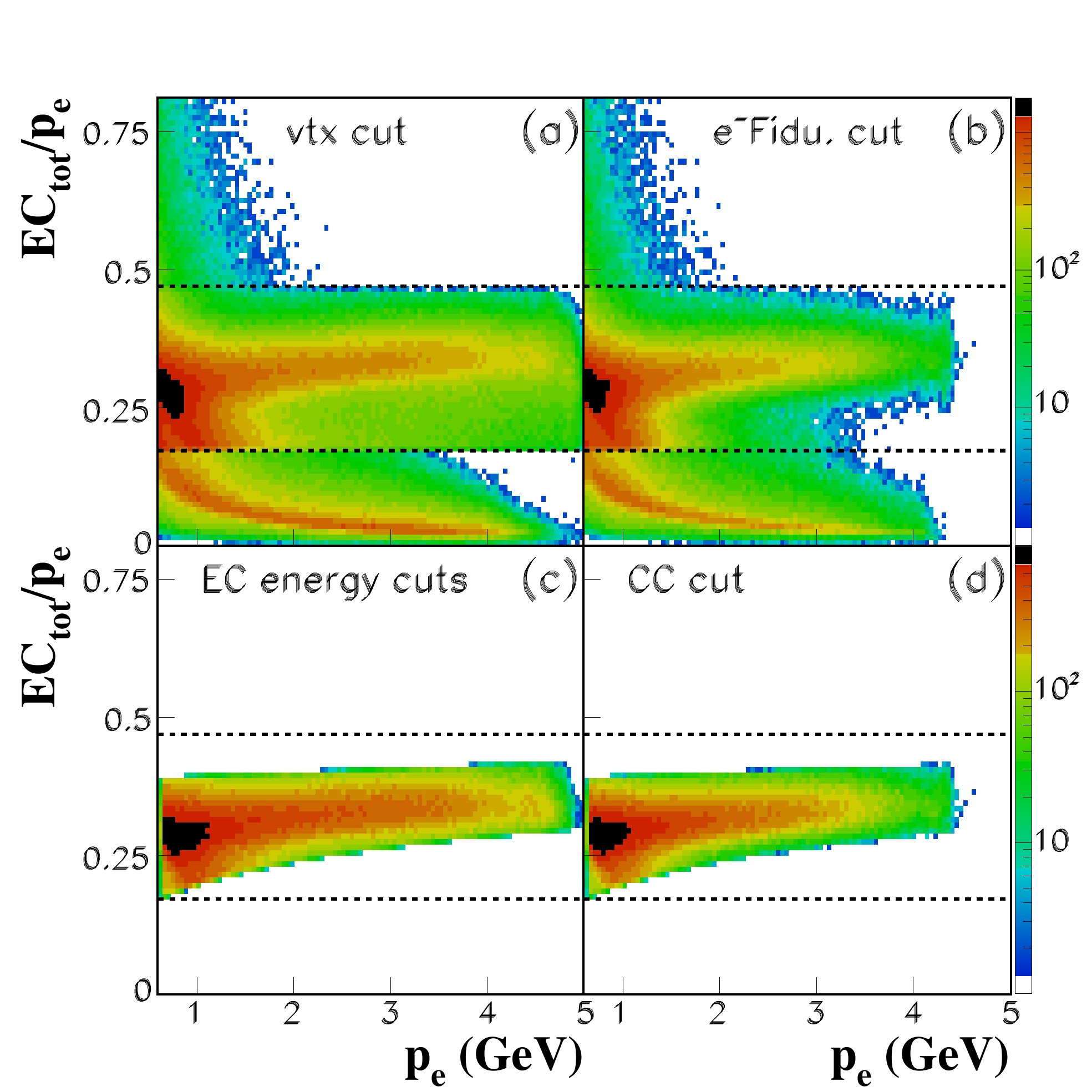}
\caption{ (Color online) Sampling fraction of energy in the EC scintillators vs. momentum of electron candidates: (a) after vertex cuts, (b) with additional electron EC fiducial cuts, (c) after EC energy cuts, and (d) after cuts on the minimum number of 
photoelectrons in the CC. The sharp transitions appearing in the top panels between the upper, middle and lower regions 
are due to the initial event selection cuts placed during the raw event "skimming". 
 \label{fig:ec_momentum}}
\end{center}
\end{figure}

In contrast to electrons, charged pions deposit energy largely though ionization, resulting in much less energy deposited in the calorimeter. Minimum ionizing pions are easily eliminated by energy cuts.  Pions which undergo hadronic interactions also deposit only a fraction of their full energy in the calorimeter volume, with more energy lost in the outer parts of the EC, while showering electrons deposit a large portion of their energy in the inner part of the calorimeter. Cuts were applied to the sampling ratio as well as to the minimum energy deposited in the EC and in the inner part ($E_{\rm inner}$). Figure~\ref{fig:ec_momentum} shows the total energy deposited in the EC scintillators versus the electron momentum before and after all cuts were applied to the sampling ratio and the total EC energy. Pions were rejected by requiring minimum deposited energy in the EC: $E_{\rm inner} > 50$~$\rm{MeV}$, and $E_{\rm total} > 140$~$\rm{MeV}$.  In addition, events were eliminated if the average number of photoelectrons recorded in the CC did not exceed 2.5 for electron candidates. Such tracks were more likely associated with negatively charged pions than with electrons. Using a Poisson distribution for the number of photoelectrons, corrections were applied for the small losses of electron events that occurred due to this cut. These corrections were done separately for all bins in $\theta_\pi$ and $\phi_\pi$ to take into account the variation of the average number of photoelectrons with kinematics.

%%%%%%%%%%%%%%%%%%%%%%%%%%%%%%%%%%%%%%%%%%%%%%%%%%%%%%%%%%%
%%%%%%%%%%%%%%%%%%%%%%%%%%%%%%%%%%%%%%%%%%%%%%%%%%%%%%%%%%%%%%%%%%%%%%

\begin{figure}[htb]
\begin{center}
	\includegraphics[angle=0,width=0.38\textwidth]{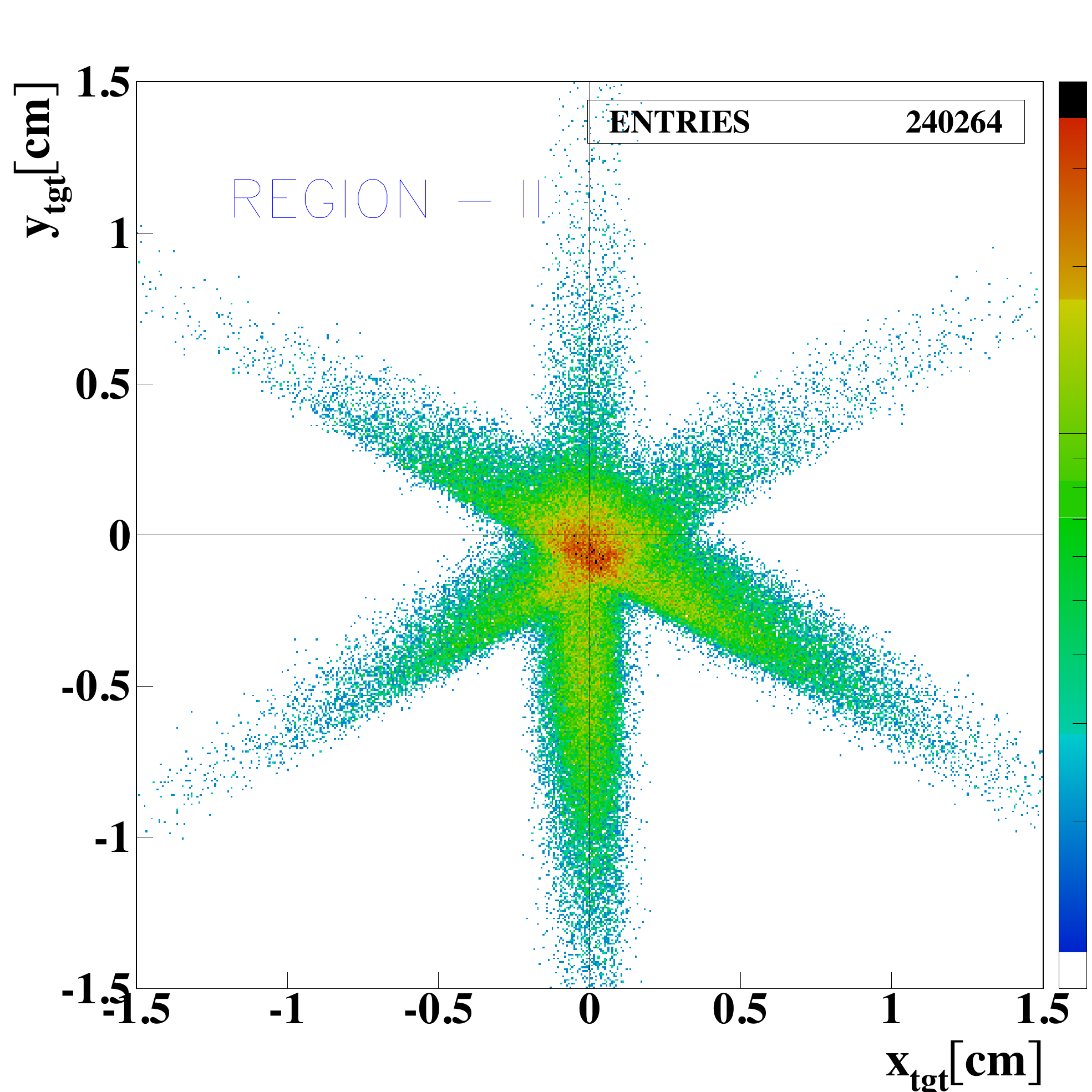}
	\includegraphics[angle=0,width=0.45\textwidth]{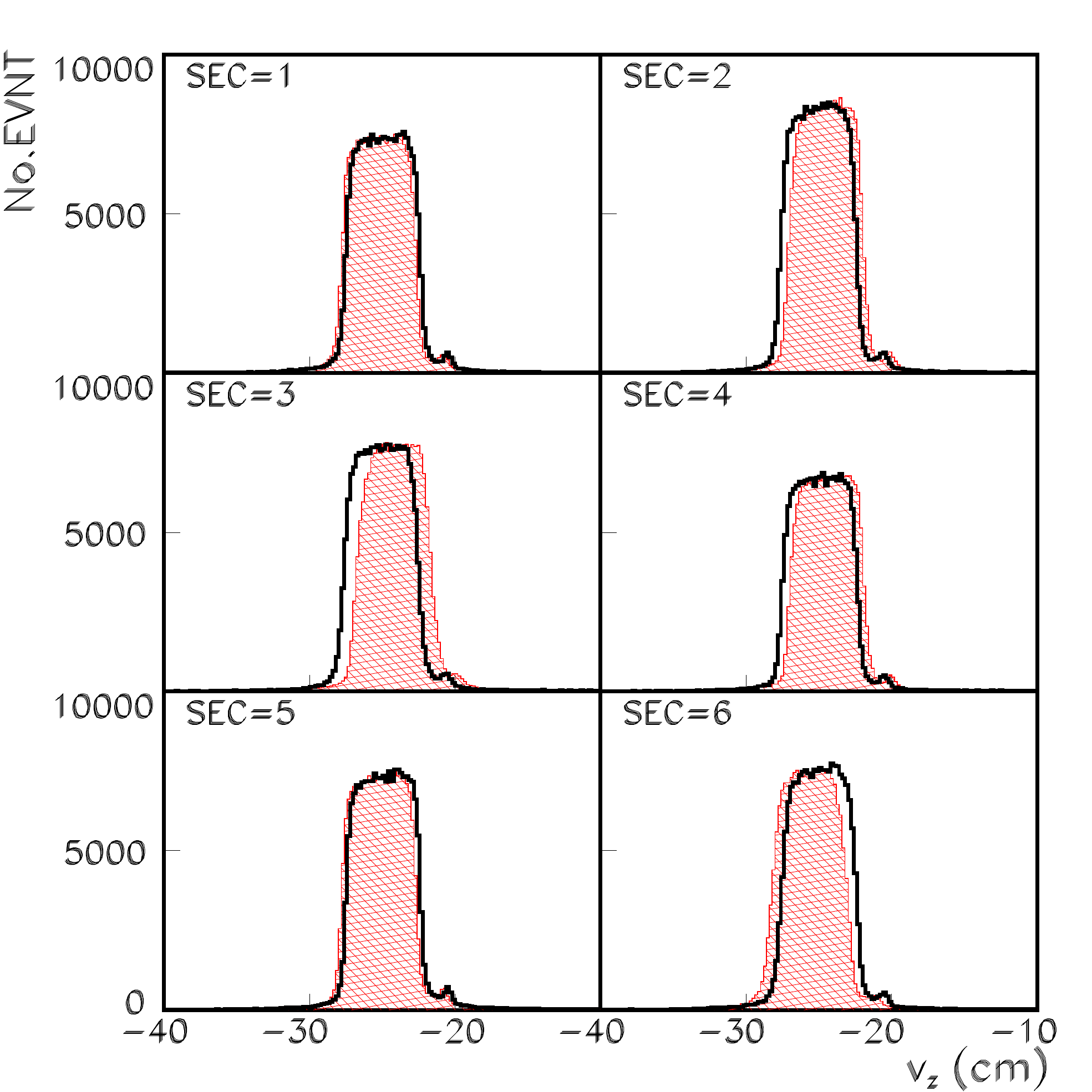}
        \caption{ (Color online) Top: Reconstructed $x$ and $y$ target position, showing a vertical offset of about -0.5 mm. The bottom panel shows the $z$-vertex before (red shaded area) and after (solid line) the beam offset in the $y$ target positions was corrected. The small enhancement near -20~cm is due to the exit window of the scattering chamber. 
          \label{fig:xyzvertex}   }
\end{center}
\end{figure}

The electron beam was centered on the hydrogen production target cell which, as can be seen in Fig.~\ref{fig:xyzvertex} (top), was located vertically about -0.5 mm relative to the CLAS center. The beam offset caused an azimuthal dependence of the reconstructed $z$-vertex $v_z$ (see Fig.~\ref{fig:xyzvertex}, bottom). After the beam offset was corrected, the azimuthal dependence of $v_{z}$ was eliminated. The small peak near $v_{z} =  -20$~cm resulted of electrons scattered from the exit window of the scattering chamber, which was located 2~cm downstream of the target cell. These events were eliminated with appropriate vertex cuts. 

After electrons were selected, the start time of the event at the vertex was determined using the reconstructed path length of the electron track and the timing in the TOF scintillator paddles. An average time resolution of $\delta T_e \approx 150~\rm{ps}$ was achieved. The vertex start time was needed to link 
the event to the beam micro-bunch that caused the interaction and to determine the velocity of the charged hadrons in the event. 
%%%%%%%%%%%%%%%%%%%%%%%%%%%%%%%%%%%%%%%%%%%%%%%%%%%%%%%%%%%%%%%%%%%%%%
%%%%%%%%%%%%%%%%%%%%%%%%%%%%%%%%%%%%%%%%%%%%%%%%%%%%%%%%%%%%%%%%%%%%%%
\subsubsection{Pion identification}        

\begin{figure*}[!htb]
\begin{center}
        \includegraphics[angle=0,width=0.47\textwidth]{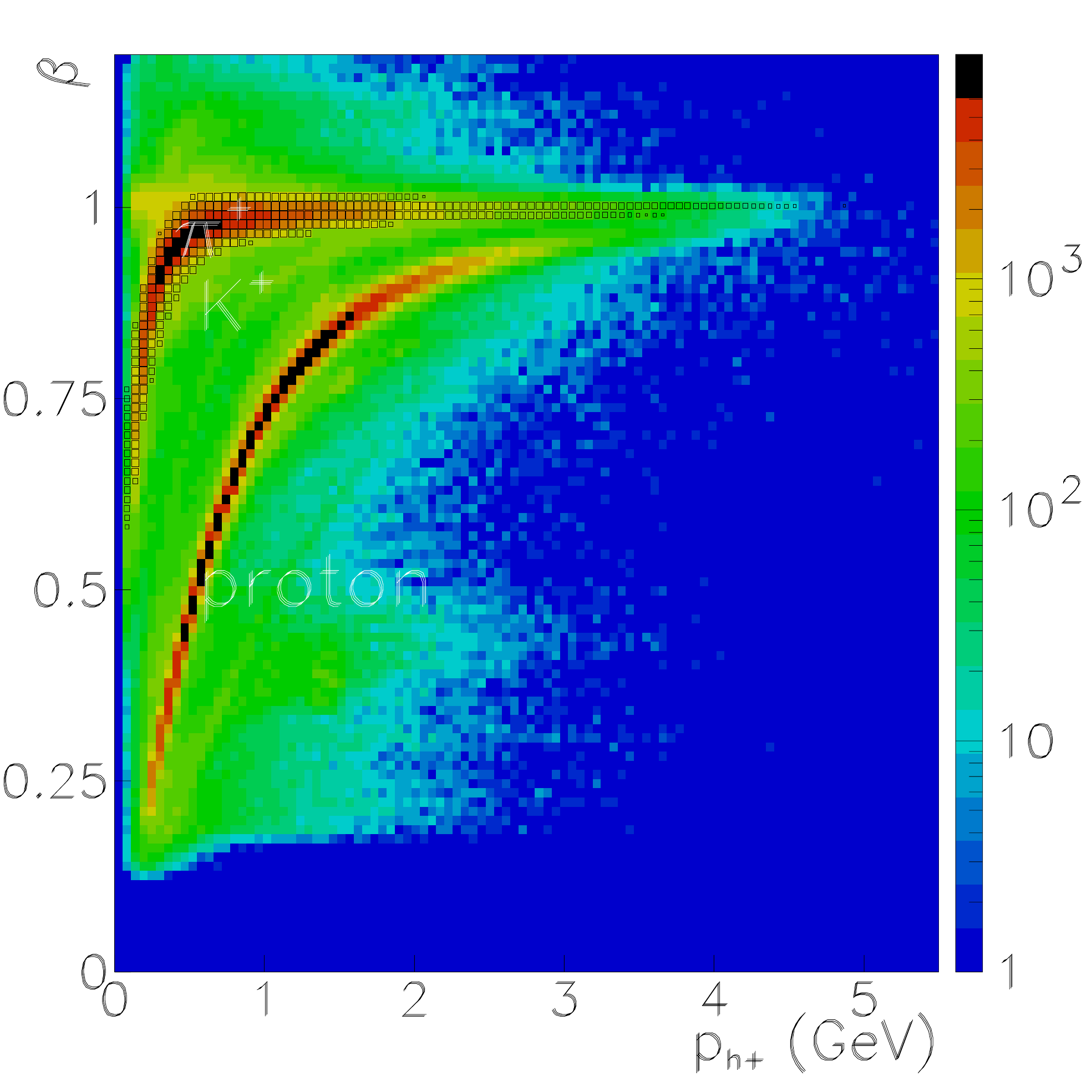}
        \includegraphics[angle=0,width=0.475\textwidth]{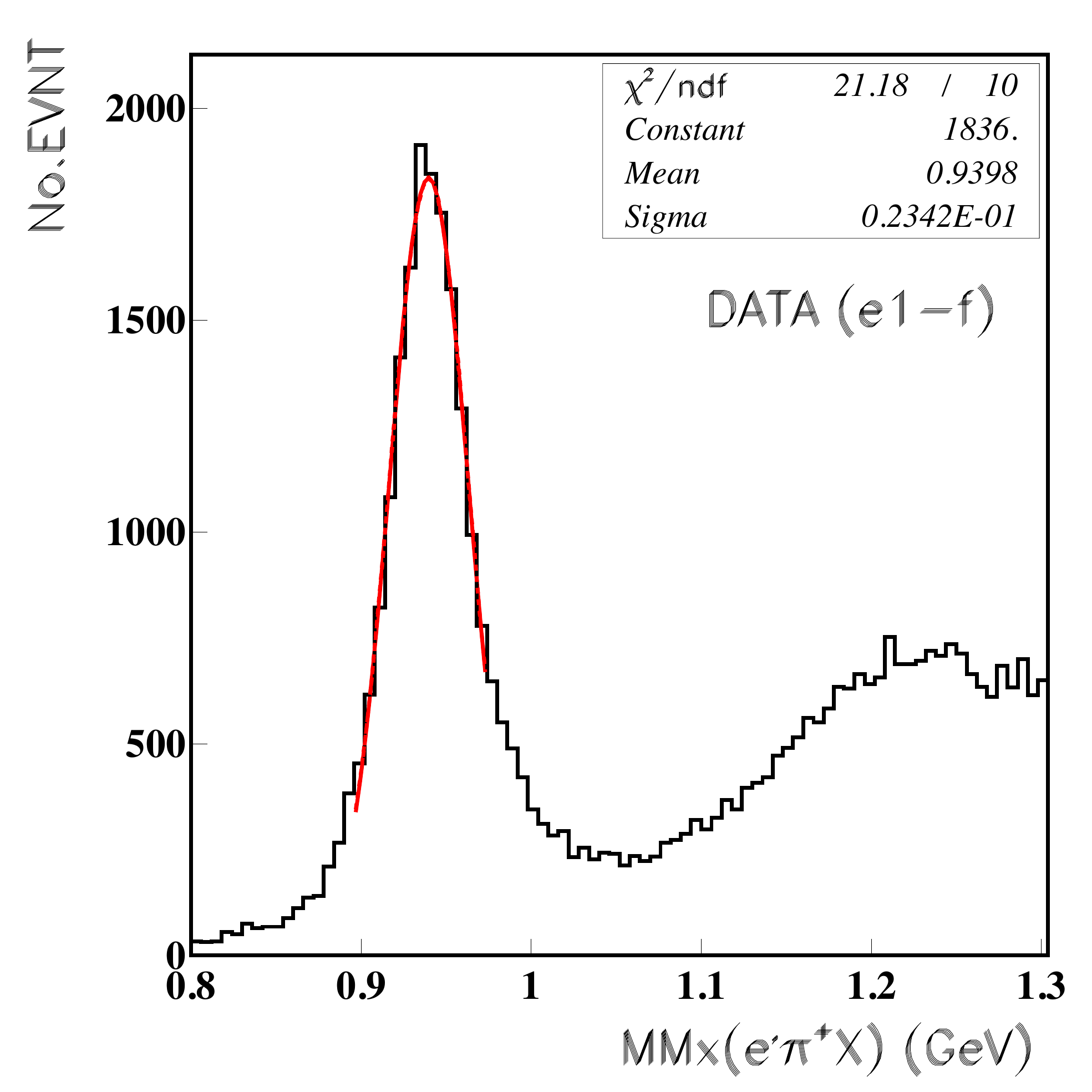}
        \caption{ (Color online) 
         \protect
          Particle velocity $\beta$ vs. momentum for positively charged hadrons (left).  
The pion and proton mass bands are clearly visible. Positively charged kaons are visible as the faint band 
between the pions and protons. The dark shaded band highlights the charged pions. 
The right panel shows the missing mass distribution ($M_X$) of $
ep \to e^\prime\pi^+X$ 
after the selection of the $\pi^+$, clearly showing the strong neutron mass peak.
         \label{fig:pion_id}
        }
\end{center}
\end{figure*}

Charged pions are identified by combining the particle velocity $\beta = v/c$, which is obtained 
from the difference of the vertex start time and the time-of-flight measurement in the TOF counters, 
with the particle momentum from tracking through the magnetic field using the CLAS drift 
chamber system. Figure~\ref{fig:pion_id} shows the charged particle $\beta$ 
versus momentum. Precise timing calibration was obtained by relating the electron timing 
to the highly stabilized radio frequency of the CEBAF accelerator. In order to isolate pions 
from protons a 3$\sigma$ cut on $\beta$ vs. $p$ was applied. Using the detected electrons 
and the isolated pions, the missing neutrons can be reconstructed through missing mass 
technique. The missing mass distribution of $ep \to e^\prime \pi^+X$ integrated over all 
kinematics is displayed in the right panel of Fig.~\ref{fig:pion_id}.  At high particle momenta 
the charged particle bands may overlap and especially kaons  may be misidentified as pions. 
These contributions lead to tails in the missing mass distributions which were estimated and 
subtracted using a procedure described in Section~\ref{sec:systematic}.
%
%%%%%%%%%%%%%%%%%%%%%%%%%%%%%%%%%%%%%%%%%%%%%%%%%%%%%%%%%%%%%%%%%%%%%%%%%

\subsection{Channel identification}
\begin{figure*}[htb]
\begin{center}
        \includegraphics[angle=0,width=0.8\textwidth]{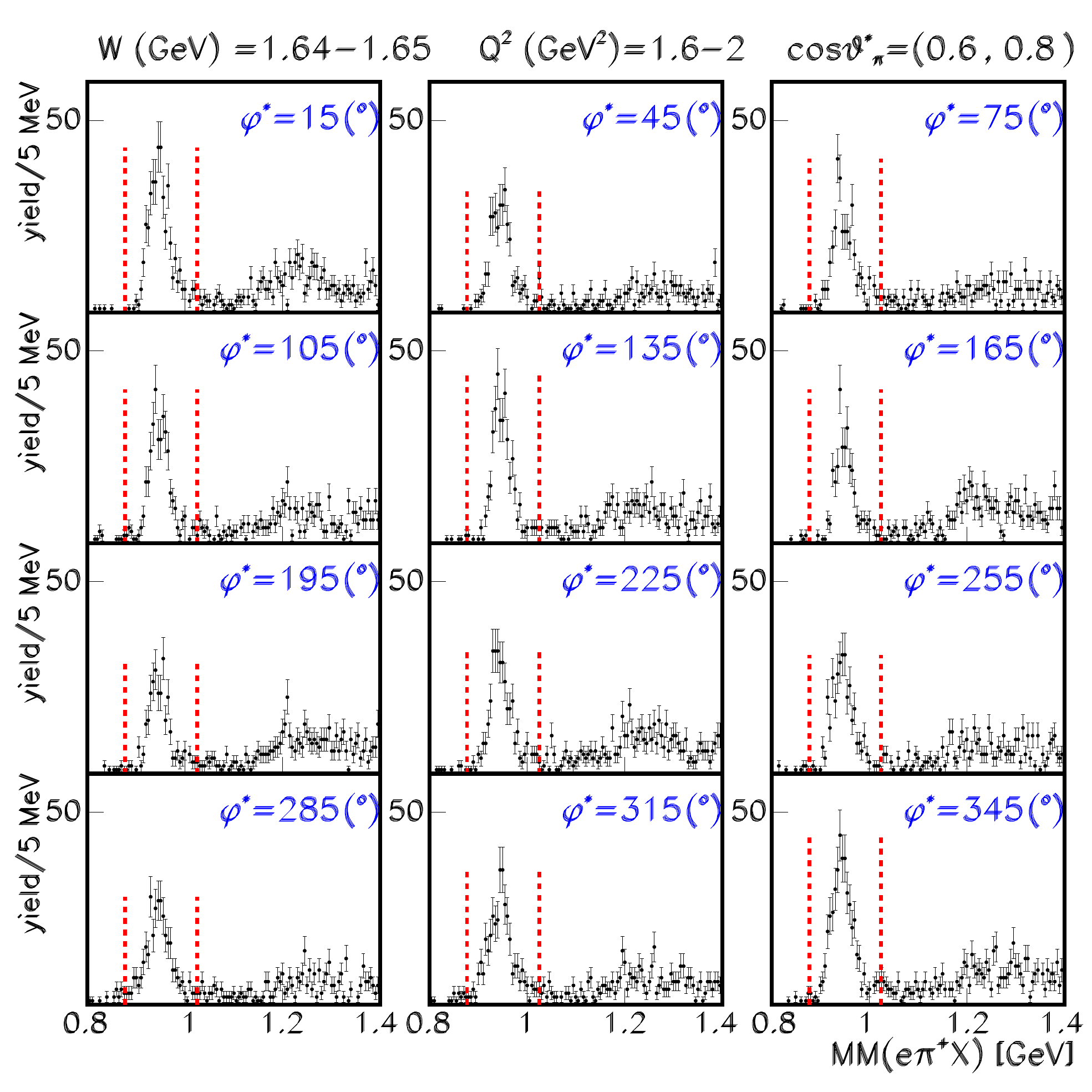}
        \caption{
        Missing mass $M_X$ distribution for $e p \to e^\prime \pi^+ X$ events for one kinematic bin in $W$, $Q^2$, and $\cos{\theta_\pi}$ for different $\phi^*$ bins.  The two vertical lines indicate the position of the event selection cuts. Background below the neutron mass peak is nearly absent as $K^+$ production near 
        the $K^+-\Lambda$ threshold is very small. 
       \label{fig:mmx_bins}
        }
\end{center}
\end{figure*}

 The final state neutron was not directly observed in this experiment. However, the four-momentum vectors of all other particles are known and four-momentum conservation and charge conservation allow the determination of the charge and the mass of the unmeasured part of the final state. The exclusive process $e p \to e^\prime \pi^+ n$ was then identified by a sharp peak in the missing mass distribution.  An example of the event distribution versus $M_X$ is shown in the right panel of Fig.~\ref{fig:pion_id}. The narrow peaks at the neutron mass indicate the exclusive process we aim to measure. The tail at the higher mass side of the neutron peak is mostly due to radiative processes. On the lower mass side of the neutron peak there are indications of some background contributions which are mostly due to kaons that are misidentified as pions in the region of higher momenta where the two particle bands shown in the left panel of Fig.~\ref{fig:pion_id} partially overlap. The background was subtracted as discussed 
 in Section~\ref{sec:systematic}. The broad enhancement near 1.2~$\rm{GeV}$ is due to the process $e p \to e^\prime \pi^+ \Delta^0(1232)$ and is not further considered. Fig.~\ref{fig:mmx_bins} shows the $M_X$ distribution versus $\phi^*_\pi$ for one specific  kinematic bin in $W$, $Q^2$, and $\cos{\theta^*_\pi}$. In order to select the exclusive process with the missing neutron in the final state, the neutron peak in each kinematical bin was fit with a Gaussian distribution, and a 3$\sigma$ cut was applied to separate the $n \pi^+$ final state from double pion production $\pi^+ (\pi N)$. This cut also eliminated some events which are part of the radiative tail for single pion production. These losses were during the extraction of the unradiated cross section. This is discussed in Section~\ref{sec:radiative_corrections}. 
%
%%%%%%%%%%%%%%%%%%%%%%%%%%%%%%%%%%%%%%%%%%%%%%%%%%%%%%%%
%%%%%%%%%%%%%%%%%%%%%%%%%%%%%%%%%%%%%%%%%%%%%%%%%%%%%%%%%%%%%%%%%%%%%%%%%%%%%%%%%%%
\subsection{Kinematic corrections}
Evidence for the need of kinematical corrections is seen in the dependence of the elastic scattering peak observed in inclusive scattering $ep \to e^\prime X$ on the azimuthal angle. This effect is most prominent at forward polar angles where the torus coils come close to each other, and is largely due to small misalignments of the torus coils resulting in a slightly asymmetric magnetic field distribution. To compensate for the small magnetic field distortions, corrections were made to the reconstructed particle momentum vector. As a first step we use the kinematically constrained elastic $ep \to e^\prime p^\prime$ process to correct for possible distortions in the reconstructed scattering angle. The proton angle was well-measured at large scattering angles where the tracking system was well aligned, and we assumed it to be accurately known, while scattered electrons were detected at small angles where the alignment of the tracking chambers was less well known, and small position shifts could result in significant shifts in the reconstructed angles. Given these conditions, the electron scattering angle could then be predicted and compared with the measured angle.  The corrections turn out to be less than 1 mrad for most of the phase space, however close to the torus coil corrections can be up to 5~mrad. 

Electron momentum corrections were derived from the difference between the predicted and measured momenta, using the corrected polar angles for elastically scattered electrons.  The magnitude of these corrections decreased to less than 0.5\% with increasing scattering angle, but could be up 1.5\% close to the torus coils. Corrections to the polar angle of the $\pi^+$ were applied using the angle corrections previously determined for electrons. The $\pi^+$ momentum was corrected by matching the observed 
missing mass $M_X$ to the neutron mass in the process $e p \to e^\prime \pi^+ X$. The exclusive process $e p \to e^\prime \pi^+ n$ was determined with an average neutron mass resolution of $\sigma_n \approx 23.4~\rm{MeV}$. 

The kinematic corrections were tested using other exclusive processes with a neutral particle in the final state, e.g. $e p \to e^\prime p \pi^0$, $e p \to e^\prime p \eta$, and $e p \to e^\prime p \omega$. In all cases, the mass of the undetected particles was reconstructed with better than 2 $\rm{MeV}$ accuracy. We take this as evidence that the kinematics of the measured particles were well determined after all corrections were applied.  

\subsection{Fiducial volumes}

The $e p \to e^\prime \pi^+ n$ reaction has been simulated in the entire phase space allowed by the incident beam energy and the CLAS acceptance. However, the CLAS acceptance is a complicated function of the kinematical variables, and there are areas, e.g. the mechanical support structure of the $\check{\rm C}$erenkov counter mirrors, and areas close to the CLAS torus coils, that are difficult to model with GSIM. To avoid the complication of edge effects, fiducial volumes with nominal full acceptance for particle detection were defined. These functions depend on azimuthal and polar angles, momentum, and charge, and are different for electrons and pions.
%%%%%%%%%%%%%%%%%%%%%%%%%%%%%%%%%%%%%%%%%%%%%%%%%%%%%%%%%%%%%%%%
\begin{figure}[thb]
\begin{center}
        \includegraphics[angle=0,width=0.470\textwidth]{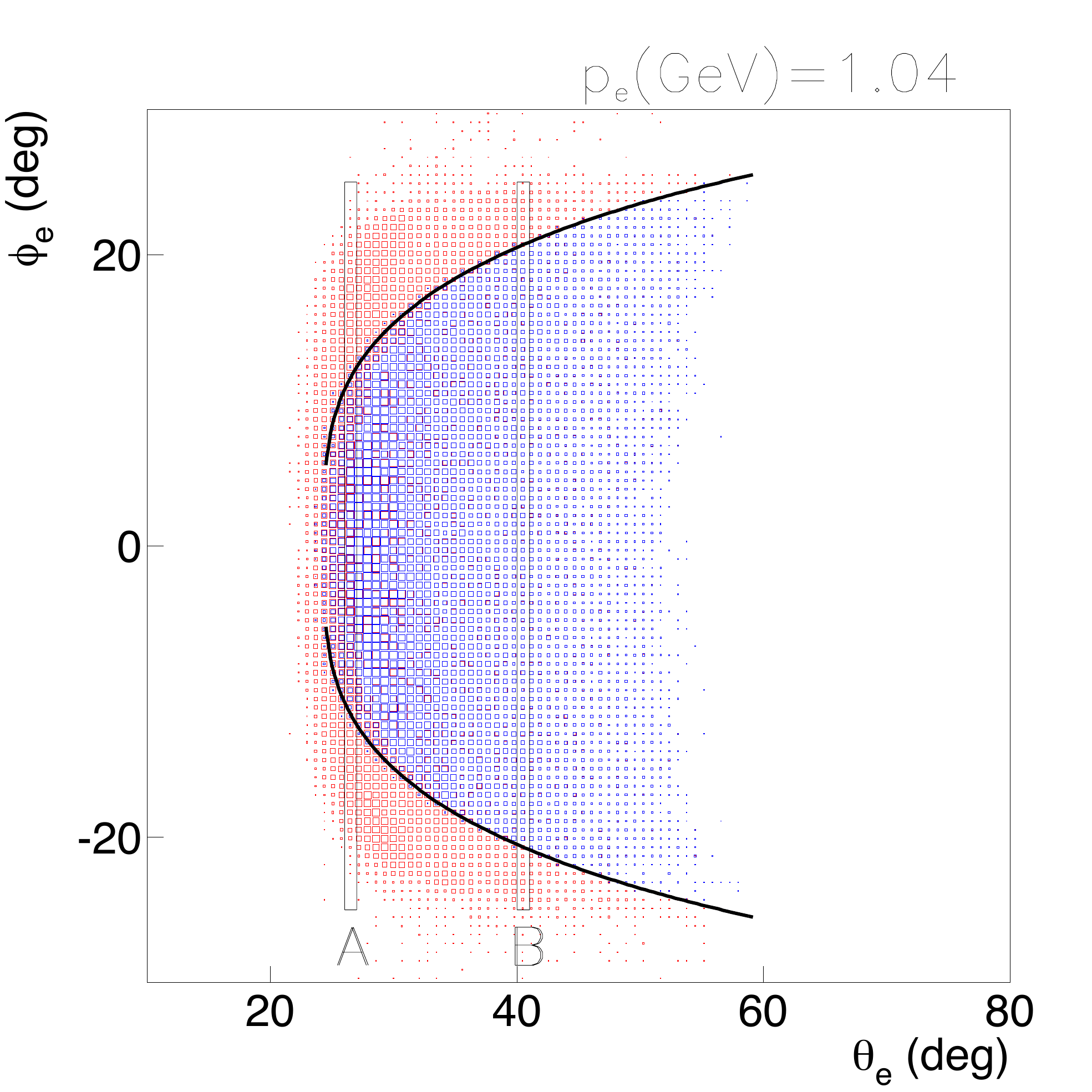}
            \includegraphics[angle=0,width=0.470\textwidth]{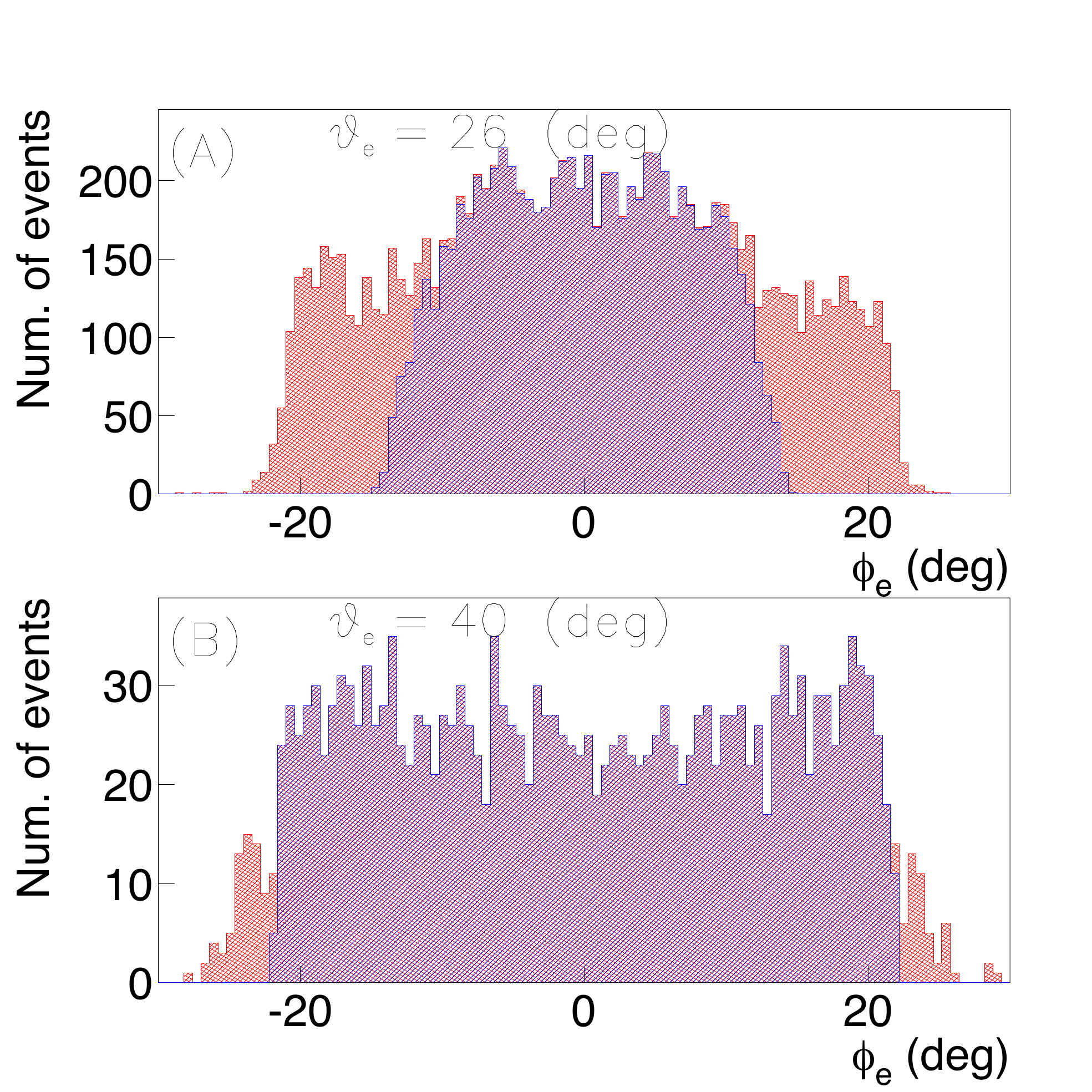}
        \caption{\protect
          (Color online) Top panel: Electron fiducial cut at $1.0 < p_e < 1.1~\rm{GeV}$ for sector 1 indicated with the outer solid lines. 
           The bottom panel 
	 shows the $\phi_e$ distributions at two values of $\theta_e$ as indicated in the top panel. The highlighted area 
	in the center indicates the selected fiducial range for the two selected polar angles.
          \label{fig:electron_fcut}
        }
\end{center}
\end{figure}
%%%%%%%%%%%%%%%%%%%%%%%%%%%%%%%%%%%%%%%%%%%%%%%%%%%%%%%%%%%%%%%%

\subsubsection{Electron fiducial volumes}
Geometrical fiducial cuts were defined to select forward regions of the detector that could be reliably simulated by the GSIM program. The $\check{\rm C}$erenkov counter efficiency has a complicated dependence on $\theta_e$ and $\phi_e$ near the acceptance edges. Fiducial volumes were defined to isolate the regions with uniform efficiency distributions. Due to the effects of the magnetic field, the angular fiducial volume also depends on the momentum of the scattered electron. The electron ($\theta_e$, $\phi_e$) distributions are shown in Fig.~\ref{fig:electron_fcut} without ({red}) and with ({blue}) fiducial cuts applied. At forward angles a rapidly varying response of the $\check{\rm C}$erenkov counters can be seen, which is due to non-uniform light collection. Applying the fiducial volume cut eliminates these regions from further analysis. 
The solid curve in Fig.~\ref{fig:electron_fcut} shows the boundary of the fiducial cut for the central momentum in that bin. Only events inside the black curve ({blue area}) were used in the analysis. In addition, a set of $\theta_e$ versus $p_e$ cuts was used to eliminate areas with reduced efficiency due to malfunctioning time-of-flight counter photomultipliers or missing drift chamber channels. The detector also contains regions with no acceptance or with low efficiency. These regions were removed as well. Holes in the acceptance are mainly due to the torus coils, and in the forward region due to the vacuum beam pipe and lead shielding surrounding the beam pipe.  
%%%%%%%%%%%%%%%%%%%%%%%%%%%%%%%%%%%%%%%%%%%%%%%%%%%%%%%%%%%%%%%%
\begin{figure}[thb]
\begin{center}
        \includegraphics[angle=0,width=0.47\textwidth]{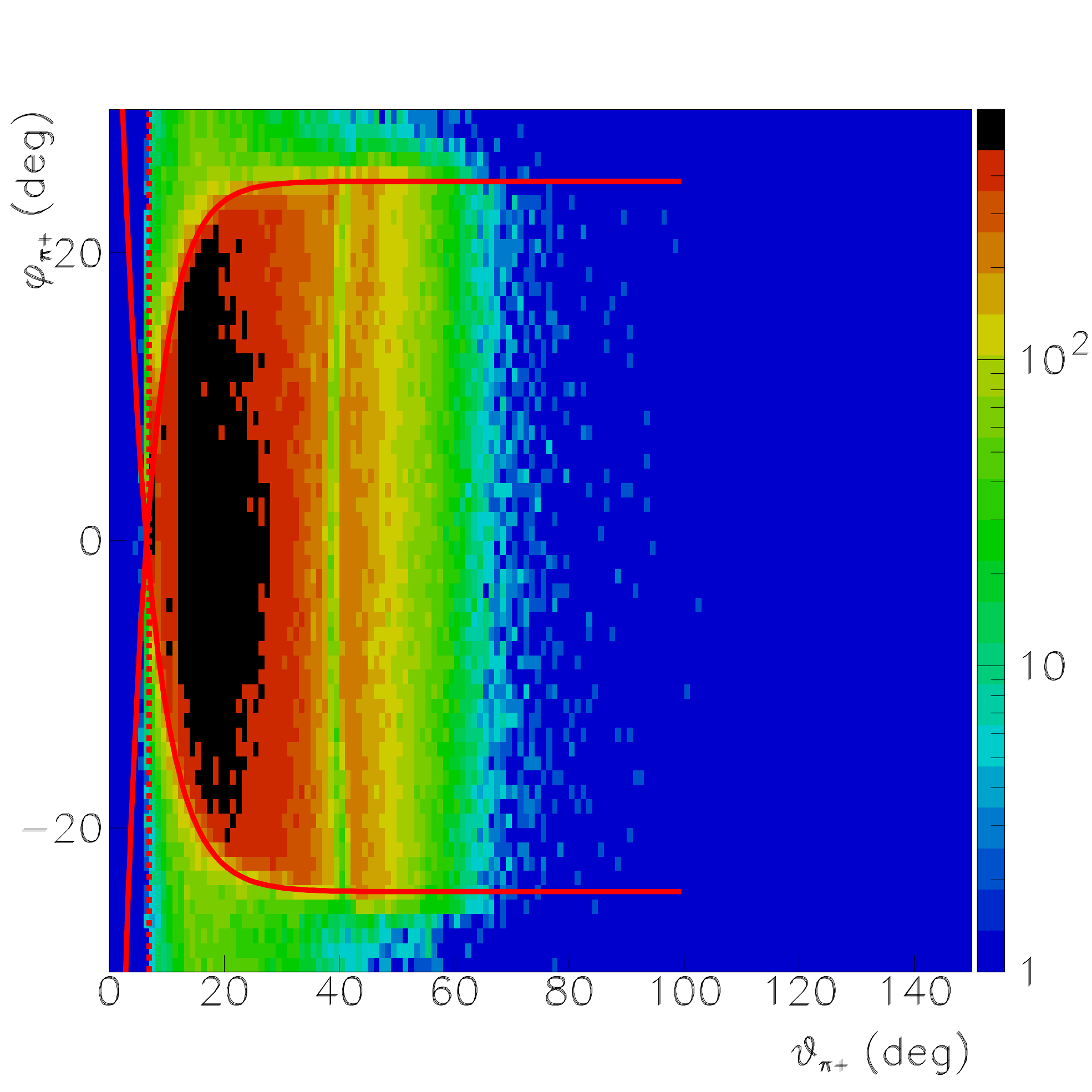}
            \includegraphics[angle=0,width=0.47\textwidth]{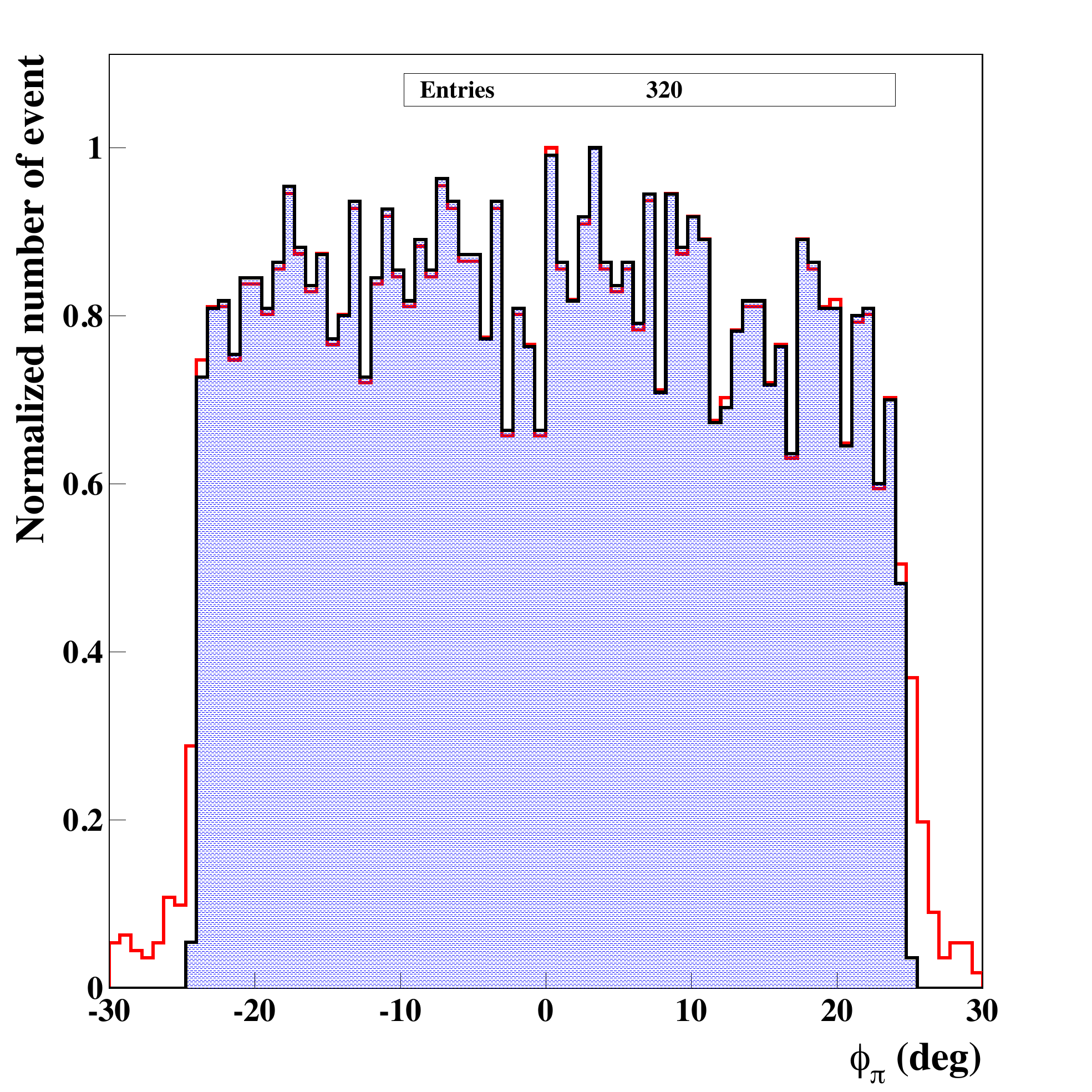}
        \caption{\protect
          (Color online) Pion fiducial cut at $0.9 < p_\pi < 1.0~\rm{GeV}$ for sector 1. The (red) solid lines in the top panel show the selected area inside the 
          $\pi^+$ azimuthal and polar angle. The histogram at the bottom
	 shows the projected $\phi_\pi$ distribution for the polar 
	 angle range $32.5^\circ < \theta_{\pi} < 34.5^\circ$  The highlighted area indicates the selected fiducial range. 
          \label{fig:pion_fcut}
        }
\end{center}
\end{figure}
%%%%%%%%%%%%%%%%%%%%%%%%%%%%%%%%%%%%%%%%%%%%%%%%%%%%%%%%%%%%%%%%
\subsubsection{Pion fiducial volumes}

The fiducial volumes for the produced $\pi^+$ are significantly different from the electron fiducial volumes. Since pion detection requires only charged particle tracking in the drift chamber system and time-of-flight measurements in the plastic scintillators, pions were detected in a much larger polar angle range from about 8$^{\circ}$ to 140$^{\circ}$.   Pion acceptance at low angles was increased by the fact that pions are outbending. An example of fiducial cuts for positive charged pions is shown in Fig.~\ref{fig:pion_fcut} 

\subsection{Kinematical binning}
\begin{figure*}[thb]
\begin{center}
        \includegraphics[angle=0,width=0.8\textwidth]{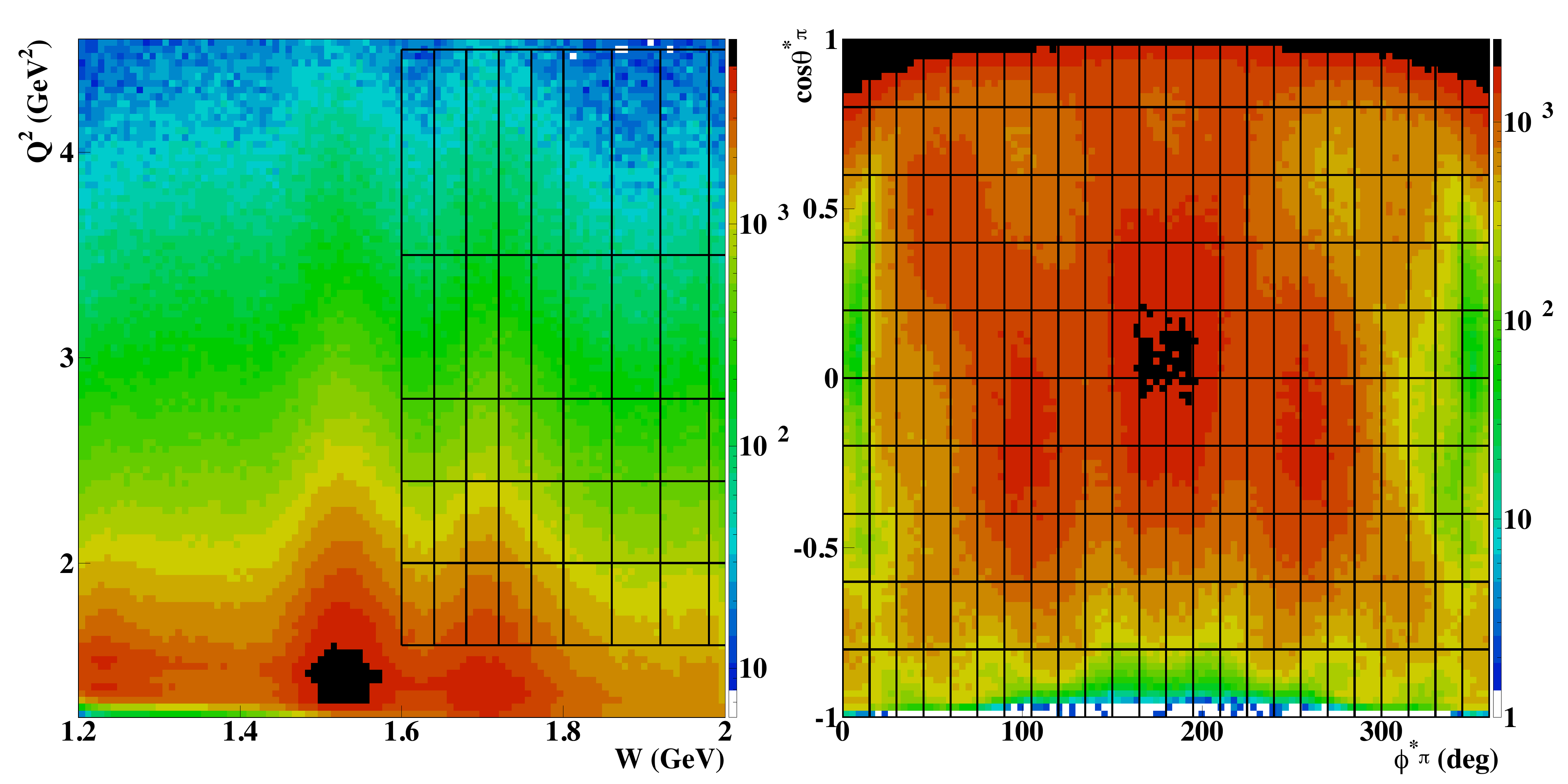}
        \caption{ (Color online) 
          Kinematic coverage in $W,~Q^2$ (left) and in $\cos{\theta^*_{\pi^*}}$,  ${\phi^*_{\pi^+}}$(right). 
	The solid lines show the bins used in most parts of the data analysis. At high $W$ and in the forward angle region $0.6 < \cos\theta^*_\pi < 1$ a finer binning in both angles was used due to the strong forward peaking of the angular distribution (not shown in the graph). 
          \label{fig:kinebin}
        }
\end{center}
\end{figure*}

The CLAS detector covers a very large kinematic range in the four CM variables $W,~Q^2,~\cos{\theta_{\pi}^*},~\phi_{\pi}$. For further analysis the data binning was matched to the underlying physics to be extracted. The study of nucleon excitations requires the analysis of the azimuthal $\phi$ dependence of the differential cross section to determine the structure functions in the differential cross section, and the analysis of the polar angle dependence to identify the partial wave contributions at a given invariant mass of the hadronic final state. The binning in the hadronic mass $W$ must accommodate variations in the cross section, taking into account the width of resonances and their threshold behavior. Table~\ref{tab:kine_range} shows the binning in $W$ and $Q^2$. The $Q^2$ binning varies as $\Delta{Q^2} = 0.2 \cdot Q^2$ to partly compensate for the rapid drop in cross section with increasing $Q^2$, while the binning in the other quantities is fixed. Figure~\ref{fig:kinebin} shows coverage in the hadronic center-of-mass angles and the binning used for the extraction of differential cross sections. As can be seen, the measurement covers nearly the entire range in $\phi^*_{\pi}$ and $\cos\theta^*_{\pi}$, with the exception of a small region near $\cos\theta^*_{\pi} = -1$, where the acceptance is significantly reduced. These regions are eliminated from the analysis by requiring a minimum acceptance for each bin. 

%%%%%%%%%%%%%%%%%%%%%%%%%%%%%%%%%%%%%%%%%%%%%%%%%%%%
\begin{table}[!htb]
\caption{Kinematical binning used in different parts of the kinematical event space to test the 
effect of the bin size. Set 1 has a fine binning in W and a coarse binning in $\phi^*_\pi$. 
Set 2 has coarse binning in $W$ and a fine binning in $\phi^*_\pi$. Set 3 covers a small part
of the polar angle range with very fine binning in $\cos\theta^*_\pi$ and in $\phi^*_\pi$.}
\begin{center}
\begin{tabular}{|c|c|c|c|c|}
Set 1 \\
\hline
Quantity & \# of Bins & Range & Bin Width \\
\hline
\hline
$W$  & 22  & $1.55 - 1.78\;\rm{GeV}$ & $10\;\rm{MeV}$ \\
\hline
$Q^2$ & 5 & $1.6 - 4.5\;\rm{GeV^2}$ & various \\
\hline
$\cos{\theta_{\pi}^*}$ & 10 & $-1.0 - 1.0$ & 0.2\\  
\hline
${\phi_{\pi}^*}$ & 12 & $0.0 - 360^o$ & $30^o$ \\
\hline \\
Set 2 \\
\hline
Quantity & \# of Bins & Range & Bin Width \\
\hline
\hline
$W$  & 9  & $1.60 - 2.0\;\rm{GeV}$ & $40\;\rm{MeV}$ \\
\hline
$Q^2$ & 5 & $1.6 - 4.5\;\rm{GeV^2}$ & various \\
\hline
$\cos{\theta_{\pi}^*}$ & 10 & $-1.0 - 1.0$ & 0.2\\  
\hline
${\phi_{\pi}^*}$ & 24 & $0.0 - 360^o$ & $15^o$ \\
\hline\\
Set 3 \\
\hline
Quantity & \# of Bins & Range & Bin Width \\
\hline
\hline
$W$  & 9  & $1.60 - 2.0\;\rm{GeV}$ & $40\;\rm{MeV}$ \\
\hline
$Q^2$ & 5 & $1.6 - 4.5\;\rm{GeV^2}$ & various \\
\hline
$\cos{\theta_{\pi}^*}$ & 10 & $0.5 - 1.0$ & 0.05\\  
\hline
${\phi_{\pi}^*}$ & 48 & $0.0 - 360^o$ & $7.5^o$ \\
\hline

\end{tabular}
\label{tab:kine_range}
\end{center}
\end{table}

%%%%%%%%%%%%%%%%%%%%%%%%%%%%%%%%%%%%%%%%%%%%%%%%%%%%%%%%
%%%%%%%%%%%%%%%%%%%%%%%%%%%%%%%%%%%%%%%%%%%%%%%%%%%%%%%

\section{Simulations \label{sec:simul}}
An essential part of the data analysis is the accurate modeling of the acceptance and event reconstruction efficiency for the process $ep \to e^\prime \pi^+n$ in the entire kinematic region accessible with CLAS.  The MAID2003 and MAID2007 physics models~\cite{MAID,maid2007} were used as event generators to populate the covered phase space as closely as possible to the measured distributions.  Nearly 200M $ep \to e^\prime \pi^+ n$ events were generated covering the measured kinematics. A GSIM Post Processor (GPP) was used to adjust the detector response such that the simulated missing mass resolution was compatible with the measured distributions. This allowed us to apply the same selection criteria for the simulated events as for the data, and gave an accurate estimate of acceptances and reconstruction efficiencies. The GPP was also used to account for missing channels in the drift chambers, and malfunctioning photomultipliers and electronics channels in the various 
detectors.  As previously discussed, cuts were applied to limit the reconstructed events to the fiducial volumes.
 
\subsection{Acceptance corrections \label{sec:accept}}  
Although the CLAS detector has a large acceptance, there are important non-uniformities and inefficiencies in some areas that need to be carefully taken into account when relating the experimentally measured yields to the differential cross sections. The complexity of the geometrical acceptance convoluted with the reconstruction efficiency that depends on all kinematical variables, prohibits an analytical parameterization of the detector response. Instead, for each of the approximately 37,000 kinematic bins in $Q^2,~W,~\cos\theta_{\pi}^*$ and $~\phi_{\pi}^*$, a single number was determined that represents the combined acceptance and efficiency for this particular bin. In addition to the acceptance corrections, the data need to be corrected for radiative effects. 
External radiation is due to the initial or the scattered electron interacting with the various material layers of the CLAS detector. This contribution was included 
in the GSIM simulation. Internal radiation corrections to the cross section are described in the next section. The number of acceptance corrected events in each bin is given by:
\begin{eqnarray}
N_{corr} = N_{exp}/Acc~~~~ Acc = \frac{REC_{RAD}}{THR_{RAD}}, 
\end{eqnarray}
where  $THR_{RAD}$ is the number of generated radiative events, $REC_{RAD}$ is the number of radiative events reconstructed in the simulation, $N_{exp}$ is the number of experimentally observed events, $Acc$ is the acceptance factor, and  $N_{corr}$ is the number of acceptance-corrected and de-radiated events. The latter includes all effects related to the detector resolution, e.g. event migration from the bin in which the event was generated to another bin where it was reconstructed. 

\begin{figure}[!thb]
\begin{center}
        \includegraphics[angle=0,width=0.45\textwidth]{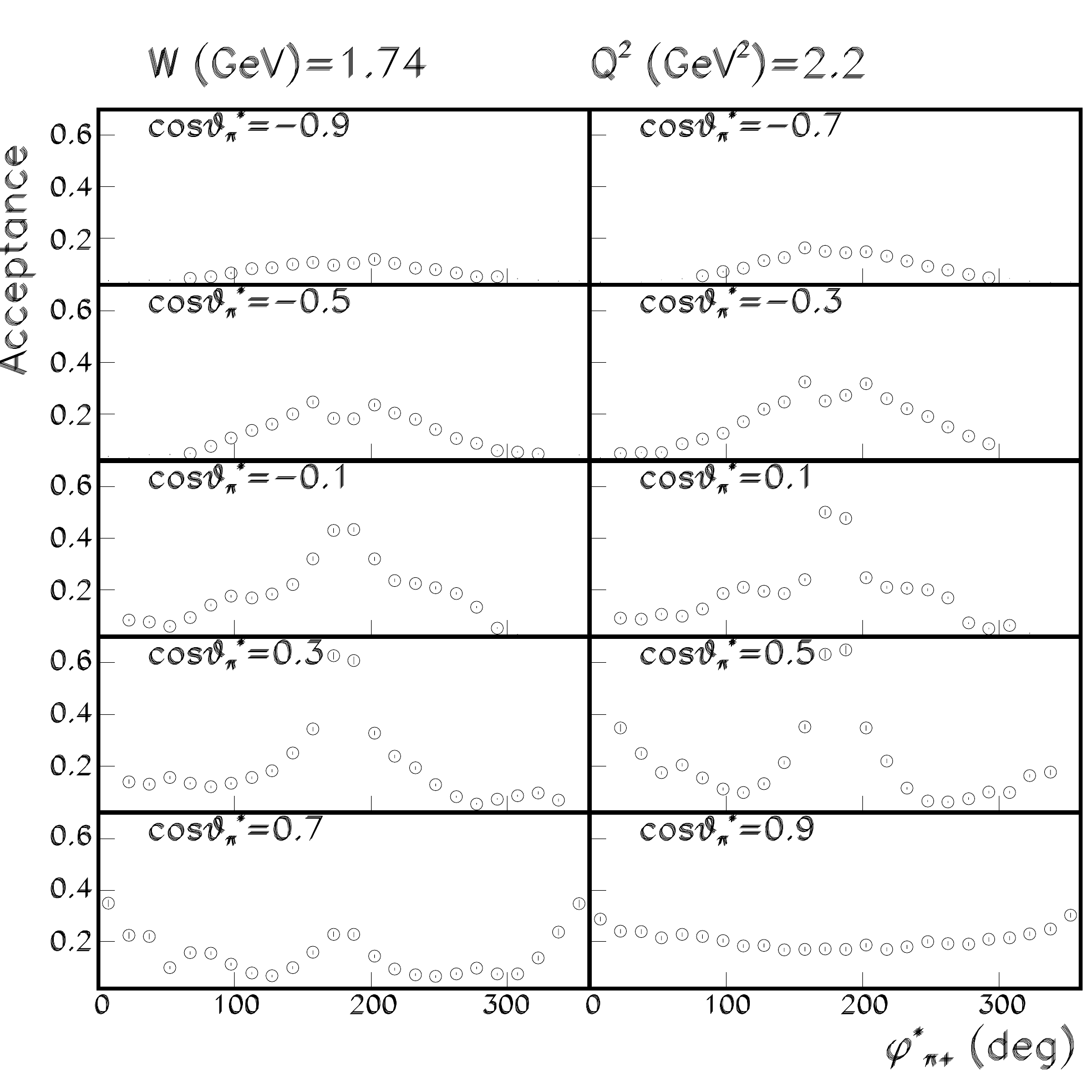}
        \caption{ \protect Acceptances for bins in azimuthal angle $\phi^*_\pi$ for several $\cos\theta^*_\pi$ bins at fixed $W=1.74$~GeV and 			$Q^2=2.2$~GeV$^2$. 
               \label{fig:accept}}
\end{center}
\end{figure}

In some regions, for example close to the torus coils, the acceptance changed rapidly with the azimuthal angle $\phi^*_{\pi}$, and could even be zero in part of the bin. To avoid inaccuracies of the acceptance calculations due to these binning effects cuts were placed to eliminate bins with acceptance of less than 2\%. This cut affected mostly the region near $\phi^*_{\pi} = 0^{\circ}$. An example of acceptance corrections is shown in Fig.~\ref{fig:accept}. The acceptance varies from a few \% to over 50\% . 

%%%%%%%%%%%%%%%%%%%%%%%%%%%%%%%%%%%%%%%%%%%%%%%%%%%%%%%%
%%%%%%%%%%%%%%%%%%%%%%%%%%%%%%%%%%%%%%%%%%%%%%%%%%%%%%%
%%%%%%%%%%%%%%%%%%%%%%%%%%%%%%%%%%%%%%%%%%%%%%%%%%%%%%%%%%%%%%%%%
\begin{figure*}[!thb]
\begin{center}
	\includegraphics[angle=0,width=0.72\textwidth]{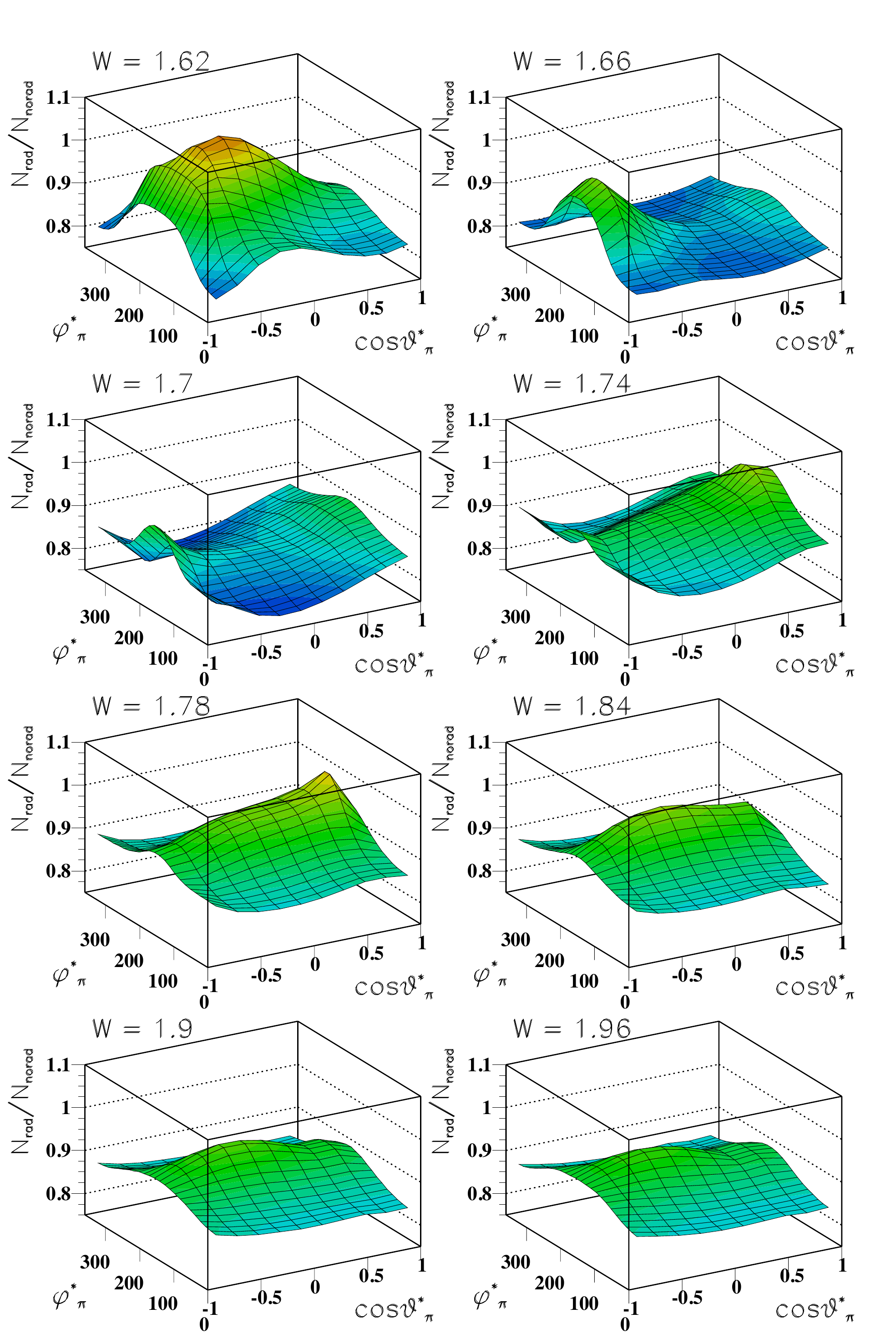}
	\caption{Examples of {\it ExcluRad} results of radiative correction factors for the pion production cross section at a specific kinematics from $W=1.62~\rm{GeV}$ to $W=1.96$~GeV and fixed $Q^2=2.6~\rm{GeV^2}$.}
	\label{fig:rad_corr_ratio}
\end{center}
\end{figure*}
%%%%%%%%%%%%%%%%%%%%%%%%%%%%%%%%%%%%%%%%%%%%%%%%%%%%%%%%%%%%%%%%%%

\subsection{Radiative corrections \label{sec:radiative_corrections}}

The often used inclusive radiative corrections cannot be applied to exclusive pion electroproduction without additional assumptions. In this analysis we have corrected the cross sections for internal radiative effects using the approach developed by A. Afanasev {\em et al.}~\cite{Afanasev:2002ee} for exclusive electroproduction of pseudoscalar mesons. This approach uses a model cross section as input, and performs an exact calculation without relying on the usual peaking approximation or the separate treatment of soft and hard photon radiation. 

Radiative processes affect the measured cross section for inclusive electron scattering. They can also modify the measured angular distributions of the hadronic final state. Therefore, a model input that closely reflects the unradiated 5-fold differential hadronic cross section is important. MAID03~\cite{MAID} was used as model input in a first step, and its parameters were adjusted subsequently to optimize the procedure. Figure~\ref{fig:rad_corr_ratio} shows as an example the $\cos\theta^*_{\pi}$ and $\phi^*_{\pi}$ dependences of the radiative correction factor 
\begin{eqnarray}
RC = \frac{\sigma_{modrad}}{\sigma_{mod}}
\end{eqnarray} 
for fixed $W$ and $Q^2$, where $\sigma_{mod}$ is the model cross section and $\sigma_{modrad}$ is the radiated model cross section.  At $Q^2 = 2.6$~GeV$^2$ and $W$ in the range 1.6 - 2.0 GeV, the radiative corrections are up to 20\% and have a non-negligible effect on the azimuthal and polar angle distributions in the hadronic center-of-mass.

\begin{figure*}[!bth]
\vspace{-5.0cm}
\begin{center}
	\includegraphics[angle=0,width=0.90\textwidth]{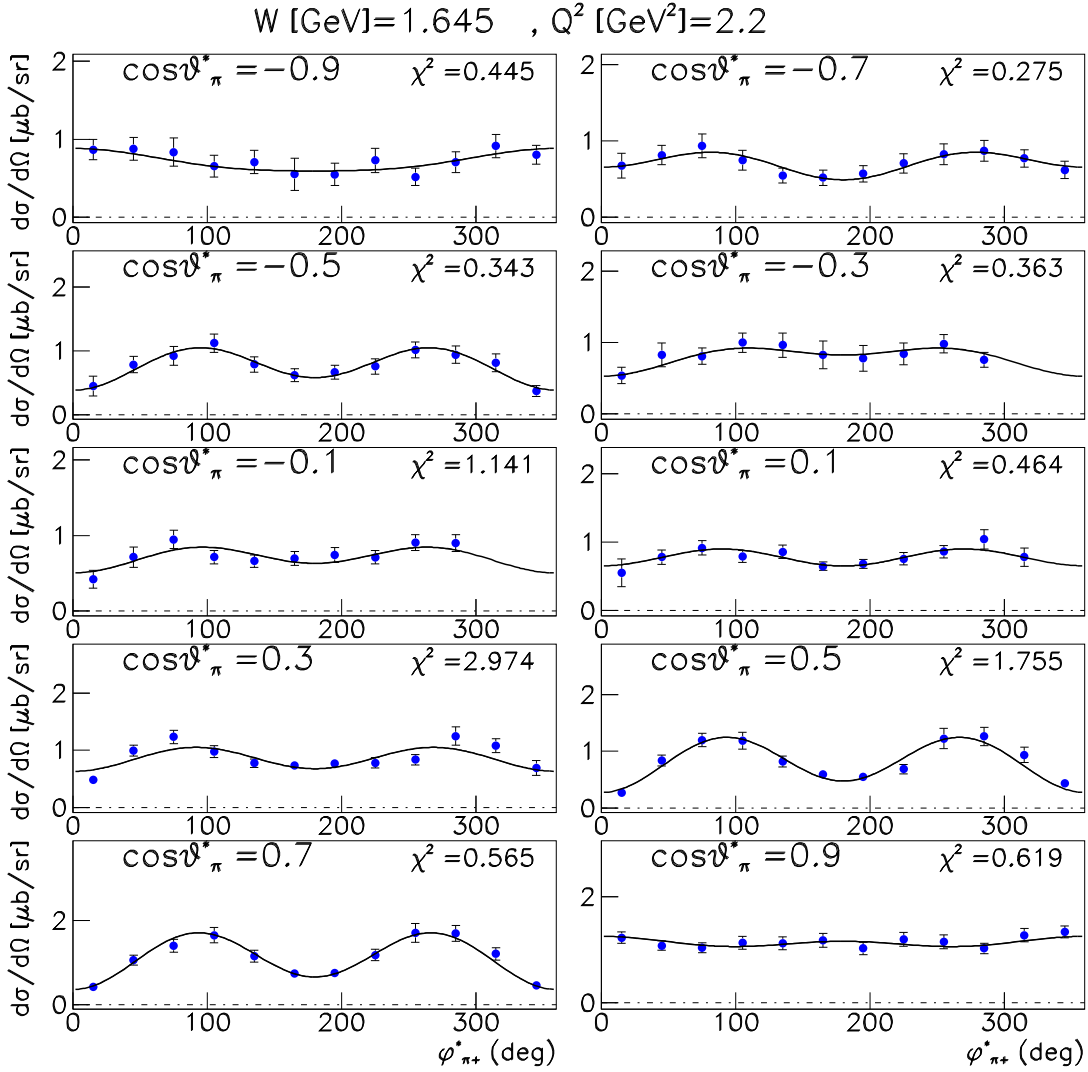}
        \caption{\protect (Color online)
          Differential cross section vs $\phi^*_{\pi}$ at $W=1.645$~GeV and $Q^2=2.2$~GeV$^2$ for different values of $\cos \theta^*_{\pi}$. The curves represent fits to the cross section using the expression given in Eq.(\ref{eq:phi-dependence}). 
          \label{fig:cs0}}
\end{center}
\end{figure*}

\begin{figure*}[!bth]
\vspace{-5.0cm}
\begin{center}
	\includegraphics[angle=0,width=0.85\textwidth]{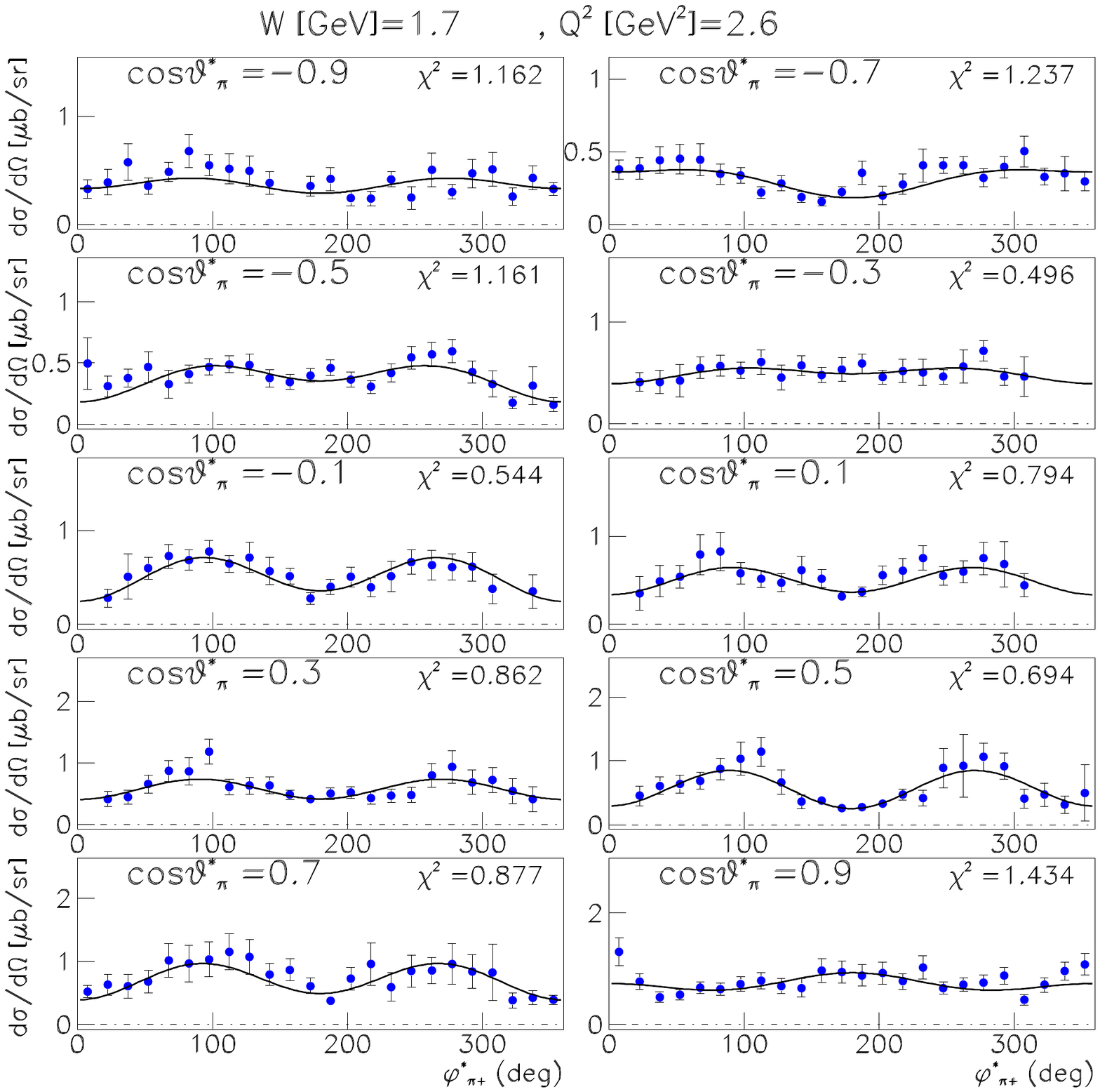}
        \caption{\protect (Color online) 
          Differential cross section vs. $\phi^*_{\pi}$ at $W=1.70$~GeV and $Q^2=2.6$~GeV$^2$ for different values of $\cos \theta^*_{\pi}$. The curves represent fits to the cross section using the expression given in Eq.(\ref{eq:phi-dependence}). 
          \label{fig:cs1}}
\end{center}
\end{figure*}

\subsection{Bin centering corrections \label{sec:bincorr}}

As the cross section can vary significantly within a given kinematics bin, the center of that bin may not coincide with the cross section weighted average within that bin. Corrections were applied to the cross section using MAID03~\cite{MAID} as a reasonable representation of these variations. The effects on the cross sections were found to be small, typically much less than $\pm 1.5\%$.

\section{Results and Discussion \label{sec:results}}

\subsection{Differential cross sections \label{sec:dcrs}}

The 5-fold differential cross section for single pion electroproduction can be written in terms of actual binned variables as in Eq.~(\ref{eq:xsec10}), using the Jacobian notation:

\begin{eqnarray}\label{eq:xsec10}
&&\frac{1}{\Gamma}\frac{d^5\sigma}{dE_f d\Omega_f d\Omega_e} = \\
&&\frac{1}{2\pi} \sum \frac{1}{L \;Acc \;\epsilon_{CC}} \frac{N f_{RC} R_{bin}}{\Delta W \;\Delta Q^2 \;\Delta \cos\theta^*_{\pi} \;\Delta \phi^*_{\pi}} \frac{d(W,Q^2)}{d(E_f,\cos\theta_e)},\nonumber
\end{eqnarray}
\noindent
where  $f_{RC}$ is the radiative correction factor and $R_{bin}$ is the bin-centering correction factor, 
$\Delta W, \Delta Q^2, \Delta\cos\theta^*_{\pi}$, and $\Delta\phi^*_{\pi}$ are the kinematic bin volumes, $L$ is the integrated luminosity, $N$ is the number of events per bin, and $\epsilon_{CC}$ is the efficiency of the $\check{\rm C}$erenkov counter. As shown in table~\ref{tab:kine_range}, different 
bin sizes were used to compute the cross section in different parts of the event space. 
The last term is the Jacobian which is defined by
\begin{equation}\label{eq:jacobian}
\frac{d(W,Q^2)}{d(E_f,\cos\theta_e)} = \frac{2 M_p \;E_i \;E_f}{W}~.
\end{equation}

Due to the large number of kinematic bins, the complete set of the resulting 37,000 differential cross section values cannot be presented in this paper. All cross sections are tabulated in the CLAS Physics Database~\cite{clas_db}. In this article we only present examples for the $\phi^*_{\pi}$ and $W$ dependences of the differential cross sections. 
From Eq.~\ref{eq:diffcrs} it is clear that the general structure of the differential cross section for single pion production with unpolarized electrons can be written as:
\begin{eqnarray}
\frac{d\sigma}{d\Omega^*_{\pi}} = A + B\cos{2\phi^*_{\pi}} + C\cos{\phi^*_{\pi}}~.
\label{eq:phi-dependence}
\end{eqnarray}
By fitting the $\phi^*_{\pi}$-dependence of the cross section we can extract the coefficients $A,~B,~C$, which depend on $Q^2$, $W$, and $\cos{\theta^*_{\pi}}$ only. They are related to the various cross section pieces as given in the following equations:
\begin{eqnarray}
 A &=& \sigma_T + \epsilon \sigma_L~,\\
 \label{eq:sigma0}
 B &=& \epsilon \sigma_{TT}~, \\
  \label{eq:sigmatt}
 C &=& \sqrt{2\epsilon(1+\epsilon)} \sigma_{LT}~.
  \label{eq:sigmalt}
\end{eqnarray}      
%%%%%%%%%%%%%%%%%%%%%%%%%%%%%%%%%%%%%%%%%%%%%%%%%%%%%%%%%%%%%
\begin{table}[!htb]
\begin{center}
\caption[Average systematic uncertainties]{Average systematic uncertainties to the differential cross sections.}
\vspace{0.5cm}
\begin{tabular}{lcll}
\hline
Source &  Contribution (\%)\\
\hline \hline
$e^-$ ID &   3.3 \\
$e^-$ fiducial cut &  2.2 \\ 
$\pi^+$ ID &  2.3 \\ 
$\pi^+$ fiducial cut &  4.5 \\ 
Missing mass selection& 2.5 \\
Vertex cut &   3.3 \\ 
Acceptance corrections&  2.1 \\  
Radiative corrections&  5.5 \\  
Binning-corrections &  1.5 \\ 
Background &   1.0 \\  
\hline
Total point-to-point& 9.5 \\
\hline\hline
Type & \\
\hline
LH2 target density &  1.0  \\
Luminosity &  3.0  \\
\hline
Total normalization &   {{3.2}} \\
\hline\hline
\label{tab:systematics}
\end{tabular}

\label{tab:sys}
\end{center}
\end{table}

Examples of the  $\phi^*_{\pi}$-dependence of the differential cross section are shown in Fig.~\ref{fig:cs0} 
and Fig.~\ref{fig:cs1} at fixed $Q^2$ and $W$ for different values of $\cos{\theta^*_{\pi}}$.
%\newpage
%%%%%%%%%%%%%%%%%%%%%%%%%%%%%%%%%%%%%%%%%
\begin{figure*}[thb]
\begin{center}
	\includegraphics[angle=0,width=0.78\textwidth]{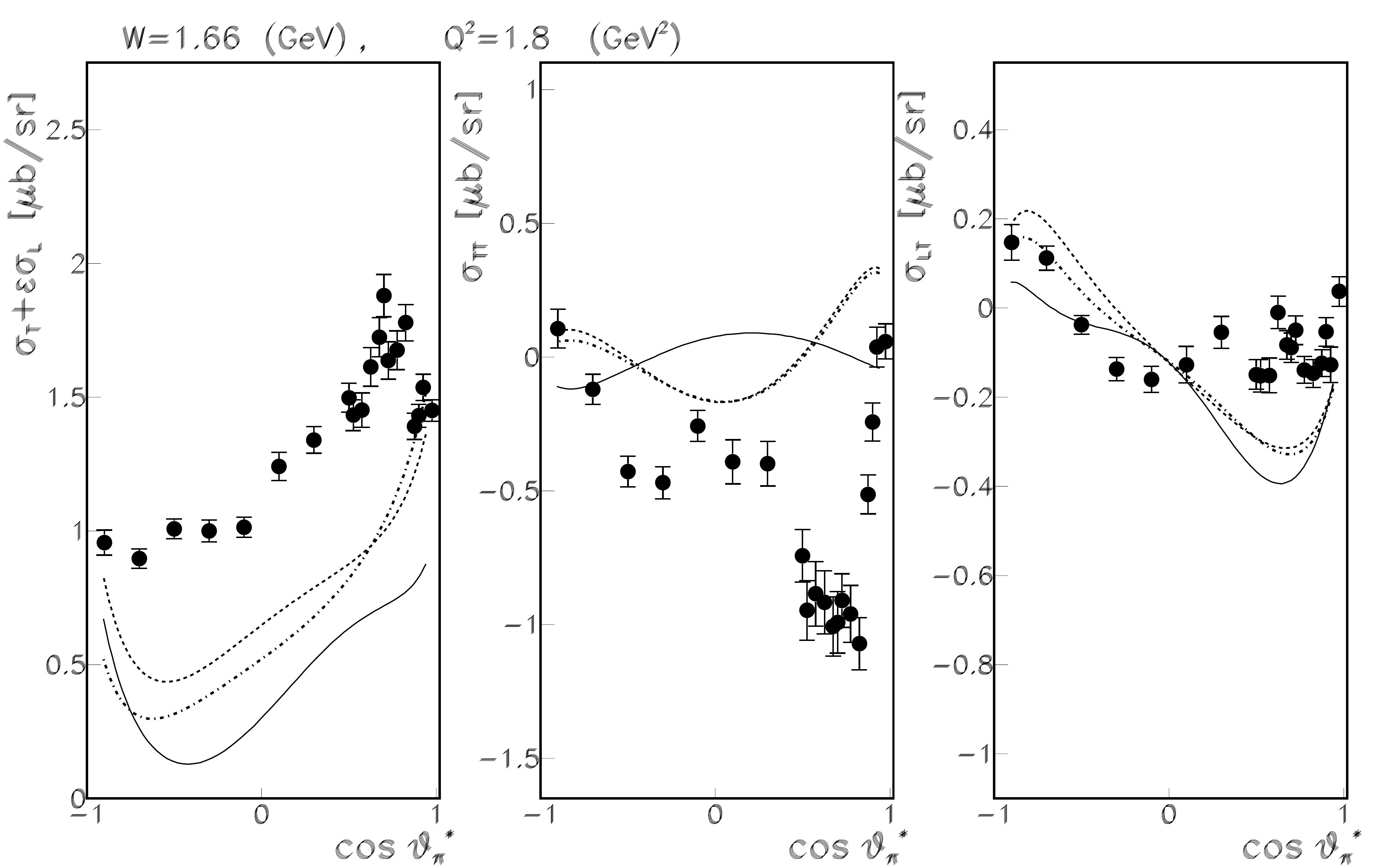}
       \includegraphics[angle=0,width=0.78\textwidth]{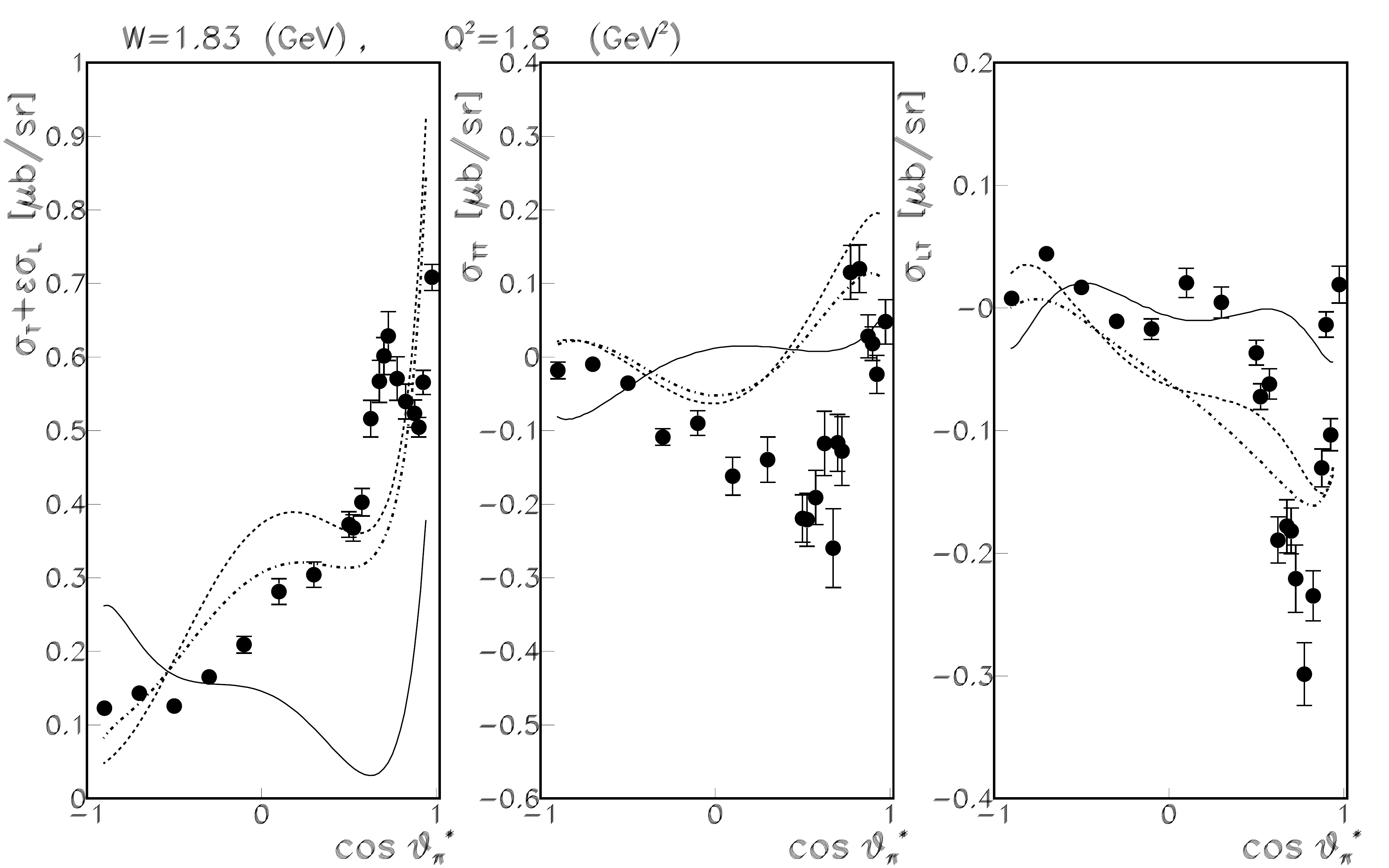}
        \caption{\protect (Color online) 
           Examples of structure functions versus $\cos\theta^*_\pi$ at fixed $Q^2$=1.8~GeV$^2$, and 
           for $W$=1.66~GeV 
           (top) and $W$=1.83~GeV (bottom). The points in the more densely populated angle range 
           of $\cos{\theta^*_\pi > 0.50}$ are from fits to cross sections measured with finer bins in $\theta^*_\pi$. The
           fine binning was needed to resolve the sharp structures seen at the forward angles. The points at  backward angles are from fits to cross sections in wider bins. The curves are projections from the dynamical models - DMT (thin-dashed), MAID2003 (dashed-dotted), MAID2007 (bold-dashed).}
          \label{fig:structure_functions}
\end{center}
\end{figure*}
%%%%%%%%%%%%%%%%%%%%%%%%%%%%%%%%%%%%%%%%%%%%%

\begin{figure}[hbt]
\begin{center}
\includegraphics[angle=0,width=0.48\textwidth]{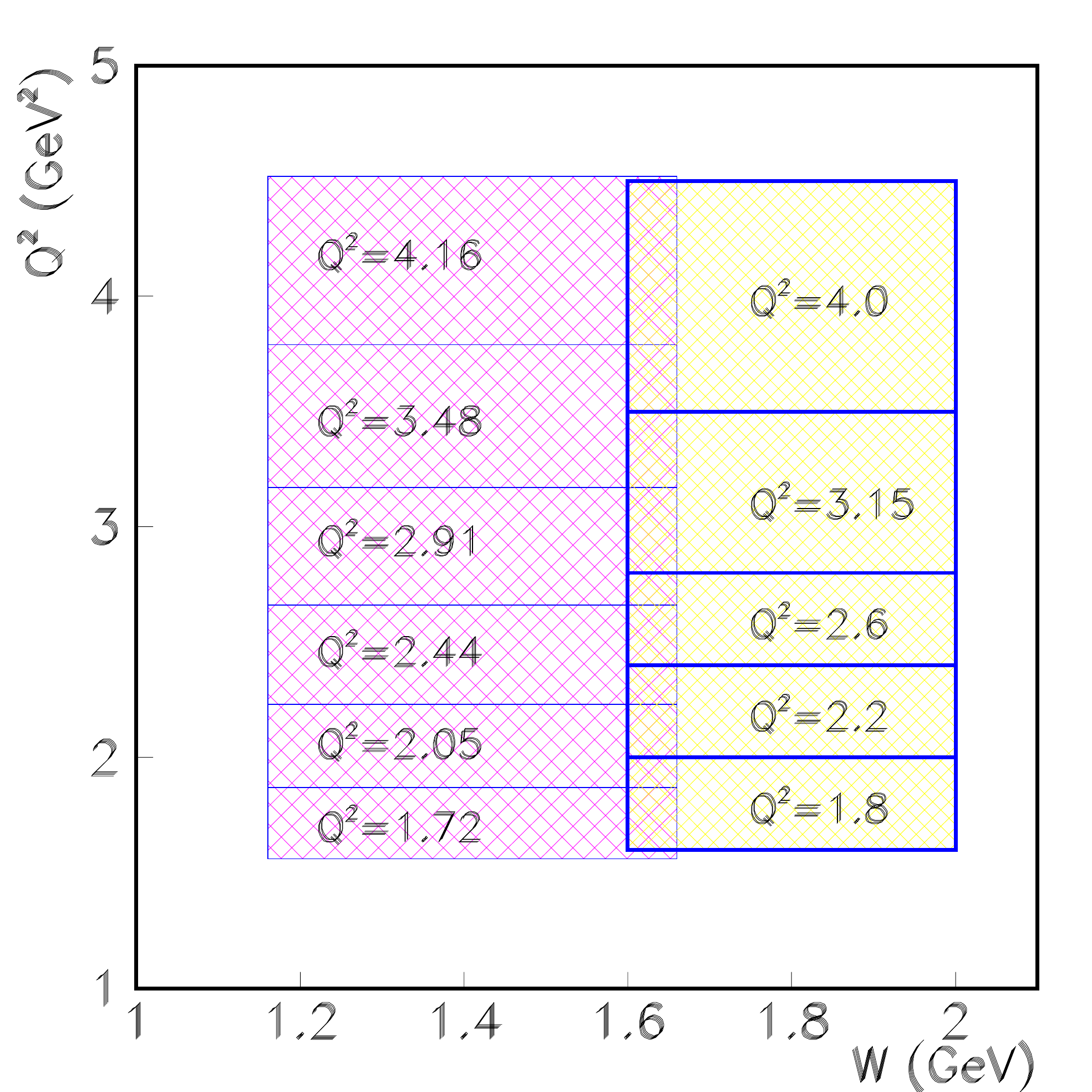}
\caption{\protect (Color online) Kinematics in $Q^2$ and $W$ for the two data sets. The data set at the lower $W$ range were published previously ~\cite{Park:2007tn}. They cover approximately the same range in $Q^2$ but are split into six bins, while the current data are binned into five $Q^2$ bins. The $W$ range of the previous measurement covered the range from pion threshold up to $W=1.69$~GeV, while the current data set covers the upper mass range from $W=1.6$~GeV to $W=2.01$~GeV.} 
\label{fig:overlap_e16-e1f}
\end{center}
\end{figure}

\begin{figure*}
\begin{center}
\includegraphics[width=14.cm]{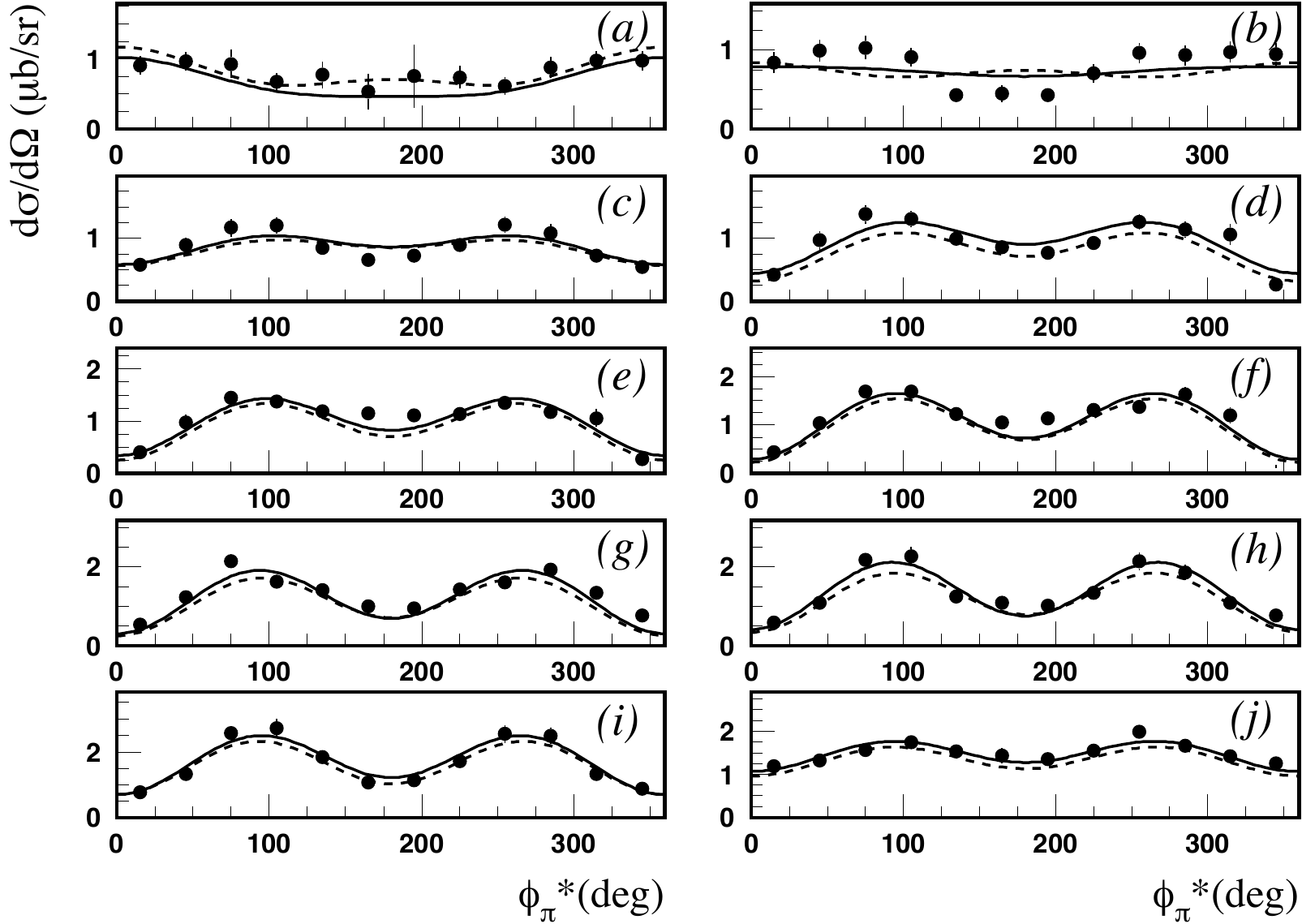}
\caption{\small Differential cross sections for the  $\gamma^* p \rightarrow n \pi^+$ reaction at $W=1.68~$GeV and $Q^2=1.8~$GeV$^2$. The panels (a,b,c,d,e,f,g,h,i,j) correspond, respectively, to $\cos \theta^*_{\pi} =-0.9,-0.7,-0.5,-0.3,-0.1,0.1,0.3,0.5,0.7,0.9$. The error bars represent the statistical and systematic uncertainties
 added in quadrature. The solid and dashed curves are, respectively, the results obtained within the UIM  and DR analyses.
\label{sec18}}
\end{center}
\end{figure*}

\begin{figure*}
\begin{center}
\includegraphics[width=14.cm]{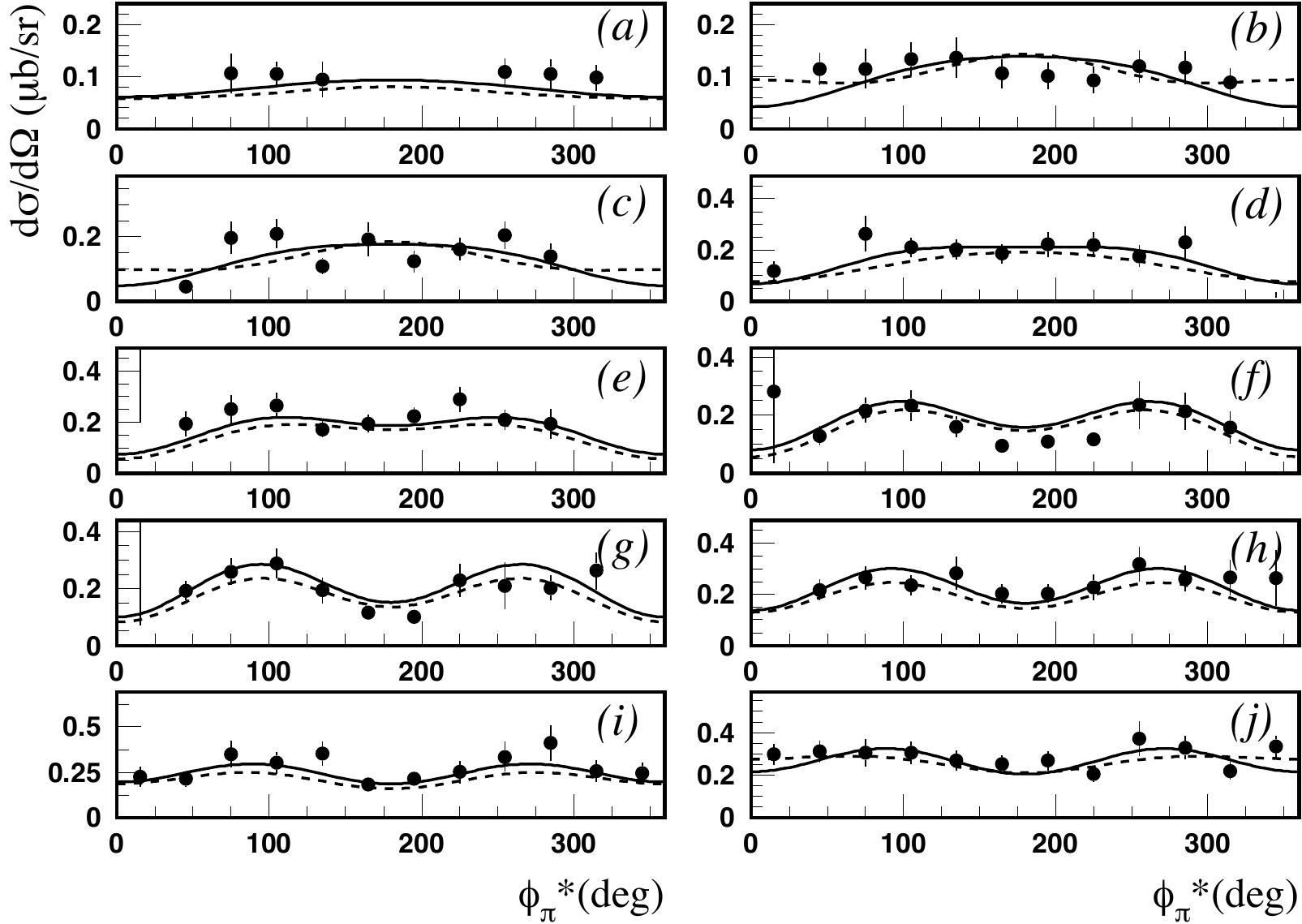}
\caption{\small Differential cross sections for the  $\gamma^* p \rightarrow~ n+\pi^+$ reaction at $W=1.68~$GeV and $Q^2=4~$GeV$^2$. The legend is as for Fig. \ref{sec18}.
\label{sec4}}
\end{center}
\end{figure*}

\subsection{Systematic uncertainties \label{sec:systematic}}
                                                                                                 
The systematic uncertainties were studied and determined with regard to 
the sensitivity of the cross section measurements to various sources of 
systematic uncertainties, e.g. by changing cut values and parameters.             

We varied the selection criteria used for the particle identification to provide 
more stringent and less stringent particle selection for both experimental
 and simulated data and then reran the complete analysis. A summary of 
 all studied sources and magnitudes of the assigned systematic uncertainties are given in 
 Table~\ref{tab:systematics}. The particle identification cuts, the vertex cuts for the electrons, 
 the fiducial cuts or the pions, the missing mass cut, and the radiative corrections are our 
 major sources of systematic uncertainty. The cuts on EC energy deposition 
 and CC amplitude for the electron, as well as the cuts on the TOF timing for the 
 pion, were varied within reasonable limits. The EC sampling fraction cut led to a 3.3\% uncertainty 
 for electron identification.
Changing the TOF $\beta$ cut for pion identification gave a 2.3\% uncertainty. 
The various cuts for reaction channel identification such as fiducial, 
missing mass, and vertex cuts produced $2.2-4.5$\%, 2.5\%, and 3.3\% systematic
uncertainties, respectively.

 The systematic uncertainty of the acceptance corrections was evaluated by comparing 
analysis results using difference versions of the MAID model. We found 
variations of about 2.0\%.
The systematic uncertainty for radiative corrections was estimated similarly 
by comparing the radiative-correction factors for different versions of MAID and 
by changing the input parameter. An average 5.5\% systematic uncertainty was 
found. 

Concerning the background subtraction procedure under the neutron missing 
mass, which could be the result from $K^+$ tracks misidentified as $\pi^+$, we assigned 
the $K^+$ mass to the identified $\pi^+$ and weighted the yields with  different 
production ratios for $K^+$ and $\pi^+$ to estimate the background.
This resulted in a 1.0\% systematic uncertainty associated with this procedure.

To take into account the model dependency of our bin-centering correction, 
we also introduce an uncertainty equal to the correction-factor itself which
 is, at the level of 1.5\% on average.

These latter systematic uncertainties were determined for each bin. Concerning
 overall scale uncertainties, the target length and density have a 1.0\% systematic 
 uncertainty and the
 integrated charge uncertainty is estimated at 3.0\%~\cite{klam2013}. The background 
from the target cell was subtracted based on the empty-target runs and amounted
 to $1.0$\% of our $e^{\prime}\pi^+n$ events. 
All other corrections were found to be less than $1.0\%$

The total systematic uncertainty was evaluated by adding 
all point-to-point systematic uncertainties in quadrature summed over all bins, 
the total systematic uncertainty is 9.5\%. In addition, 
the normalization uncertainty is approximately 3.2\%. 
Table~\ref{tab:sys} summarizes the main systematic uncertainties in this
 analysis averaged over all the accessible kinematic bins. We want to emphasize 
 that systematic uncertainties have been evaluated for each of the 37,000 cross sections.
 Their magnitudes vary significantly over the full ranges in $Q^2$, $W$, $\cos\theta^*_\pi$, and
 $\phi^*_\pi$. They are included in the CLAS Physics Database~\cite{clas_db}. 
 The numbers given in table~\ref{tab:sys} can therefore only provide a global picture of 
 their magnitudes.   
%

%%%%%%%%%%%%%%%%%%%%%%%%%%%%%%%%%%%%%%%%%%%%%%

\subsection{Structure functions \label{sec:structure_functions}}
The fit of the differential cross sections with the expression of~(\ref{eq:phi-dependence}) yields the 3 terms 
$\sigma_T + \epsilon \sigma_L$, $\epsilon \sigma_{TT}$ and $\sqrt{2\epsilon(1+\epsilon)}\sigma_{LT}$, 
with $\epsilon$ depending on the electron kinematics; the structure functions 
$\sigma_T$, $\sigma_L$, $\sigma_{TT}$, and $\sigma_{LT}$ are functions of $W$, $Q^2$, and $\cos\theta^*_\pi$. 
Note that the measurement was done at a fixed electron beam energy, thus the terms $\sigma_T$ and $\sigma_L$ cannot be separated. 
The $\cos\theta^*_\pi$ distribution is of particular interest at fixed $W$ and $Q^2$ as it represents 
the partial wave content and thus reflects sensitivity to $s$-channel resonance excitations, 
as well as interferences of the complex amplitudes. Examples of the $\cos\theta^*_{\pi}$ dependence of the 
extracted structure functions are shown 
in Fig.~\ref{fig:structure_functions}. The data on $\sigma_T + \epsilon \sigma_L$  show a strong forward peaking, 
which is related to the pion pole. 
We remark that the MAID curves were based on parameterizations of background and resonance 
contributions from fits to previous data and are therefore not considered model predictions. 
The discrepancy with the new data then indicates that the parameterizations used do not fully capture the
background and resonance contributions of the new data.  
In the following section we discuss global fits to the differential cross sections to obtain improved 
information about the resonance amplitudes underlying the cross section data.  
%%%%%%%%%%%%%%%%%%%%%%%%%%%%%%%%%%%%%%%%%%%%%%%  

\section{Extraction of resonance electrocouplings \label{sec:electrocouplings }}

In this section we present the results obtained in the analysis of the data within the Unitary Isobar 
Model (UIM) and the fixed-t Dispersion Relations (DR) approach. To provide further constraints in the analysis, we have combined the data reported in the present paper with the earlier CLAS data~\cite{Park:2007tn} on the cross sections and longitudinally polarized beam asymmetries in $\pi^+$ electroproduction on protons in the lower mass range $1.15 \leq W \leq 1.69~$GeV and at values of $Q^2$ that are close to those used in the current data set. The kinematics of the two data sets are shown in  Fig.~\ref{fig:overlap_e16-e1f}. 

The complete data sets consist of the present data at $Q^2=1.8, 2.2, 2.6, 3.15, 4.0~$GeV$^2$ and the corresponding data at  $Q^2=1.72, 2.05, 2.44, 2.91, 3.48, 4.16~$GeV$^2$ Ref.\cite{Park:2007tn}. When combining the two data sets from the different measurements we use the data with $Q^2$ values that are closest to each other. From the six 
$Q^2$ values of the previous measurements we do not use the  data at $Q^2 = 3.48$~GeV$^2$.    

The employed approaches of UIM and DR have been described in detail in Refs. \cite{Azn2003,Azn2009} and have been used successfully in Refs. 
\cite{Azn2009,Azn065,Azn2005} for the analyses of pion-electroproduction data in a wide range of $Q^2$ from $0.16$ to $6~$GeV$^2$. 

The UIM  \cite{Azn2003,Azn2009} has been developed on the basis of MAID~\cite{MAID}. At the values of $Q^2$ under investigation, the background of the UIM \cite{Azn2003,Azn2009} is built from the nucleon exchanges in the $s$-and $u$-channels and $t$-channel $\pi$, $\rho$, and $\omega$ exchanges. This background is unitarized via unitarization of the multipole amplitudes in the $K$-matrix approximation. The resonance contributions are parametrized in the unified Breit-Wigner form with energy-dependent widths.

The DR approach \cite{Azn2003,Azn2009} is based on fixed-$t$ dispersion relations for the invariant amplitudes. They relate the real parts of the amplitudes to the Born terms ($s$- and $u$-channel nucleon and $t$-channel $\pi$ exchanges) and the integral over the imaginary parts of the amplitudes. Taking into account the isotopic structure, there are 18 invariant amplitudes that describe $\pi$ electroproduction on nucleons~\footnote{For all these amplitudes, except one ($B_3^{(-)}$ in the notations of Refs. \cite{Azn2003,Azn2009}), unsubtracted DR can be written.
For $B_3^{(-)}$, the subtraction is necessary. At the values of $Q^2$ under investigation, the subtraction was found empirically in Ref. \cite{Azn2009} from the description of the data \cite{Park:2007tn}. This subtraction is also employed in the present analysis.}.

In Ref. \cite{Azn2003}, arguments were presented and discussed in detail which show that in $\pi$ electroproduction on nucleons, DR can be reliably used at $W \leq 1.8~$GeV. The same conclusion was made in early applications  of DR (see, for example, Ref. \cite{Crawford}). Therefore, in our DR analysis, the
energy region is restricted by the first, second, and third resonance regions.

Both global fits, using the UIM and the DR approach, give equivalent descriptions of the differential cross sections. This is also demonstrated in Table \ref{pip_data} in terms of the overall $\chi^2$ for the fits, 
and shown in Fig.~\ref{sec18} and Fig.~\ref{sec4}.  

\begin{table}
\begin{center}
\begin{tabular}{|cccccc|}
\hline
&&Number of&&$\chi^2/N$&\\
$Q^2$&$W$&data points&&&\\
(GeV$^2$)&(GeV)&($N$)&UIM&${}$&DR\\
\hline
1.72&1.15-1.69&3530&2.7&&2.9\\
1.8&1.6-2.01&8271&2.4&&\\
&1.6-1.8&5602&2.3&&2.4\\
2.05&1.15-1.69&5123&2.3&&2.5\\
2.2&1.6-2.01&8140&2.2&&\\
&1.6-1.8&5539&2.3&&2.3\\
2.44&1.15-1.69&5452&2.0&&2.3\\
2.6&1.6-2.01&7819&1.7&&\\
&1.6-1.8&5373&2.0&&2.2\\
2.91&1.15-1.69&5484&2.1&&2.3\\
3.15&1.6-2.01&7507&1.8&&\\
&1.6-1.8&5333&2.1&&2.0\\
4.16&1.15-1.69&5778&1.2&&1.3\\
4.0&1.6-2.01&5543&1.3&&\\
&1.6-1.8&4410&1.5&&1.6\\
\hline
\end{tabular}
\caption{\label{pip_data} The values of $\chi^2$ for the $\gamma^* p \rightarrow \pi^+ n$ cross sections
obtained in the analyses within the UIM and DR approaches. The data at $Q^2=1.8,2.2,2.6,3.15,4$~GeV$^2$ and $Q^2=1.72,2.05,2.44,2.91,4.16$~GeV$^2$ are, respectively, from the present work and 
Ref. \cite{Park:2007tn}.}
\end{center}
\end{table}
\begin{table}[t]
\begin{tabular}{|l|ccc|}
\hline
&&&\\
$~~Q^2$&$A_{1/2}$&$A_{3/2}$&$S_{1/2}$\\
(GeV$^2$)&&&\\
&&&\\
\hline
   1.8&$  13.6\pm 0.9\pm 0.7$
&$  -1.0\pm 1.0\pm 2.3$&$ -3.1\pm 1.2\pm 1.7$\\
   2.2&$  11.6\pm 0.8\pm 0.5$
&$  -2.1\pm 1.5\pm 1.1$&$ -2.1\pm 1.2\pm 0.8$\\
   2.6&$  7.6\pm 1.4\pm 0.6$
&$  -3.2\pm 1.5\pm 1.2$&$ -2.0\pm 1.3\pm 1.1$\\
   3.15&$ 5.7\pm 1.4\pm 1.3$
&$  -2.2\pm 1.3\pm 1.7$&$ -2.5\pm 1.1\pm 1.9$\\
   4.0&$ 2.4\pm 1.2\pm 1.3$
&$   -1.4\pm 1.3\pm 1.7$&$  -1.2\pm 1.3\pm 2.3$\\
\hline
\end{tabular}
\caption{\label{d15_av}
The average values of the
$\gamma^* p \rightarrow~N(1675){5\over 2}^-$
helicity amplitudes
found using UIM and DR (in units of $10^{-3}$GeV$^{-1/2}$).
The first and second uncertainties are, respectively, the statistical uncertainty from the fit 
and the model uncertainty discussed in the text.
The amplitudes are extracted using the following mass, width, and $\pi N$ branching
ratio of the resonance: 
$M=1.675~$GeV, $\Gamma=0.15~$GeV, and $\beta_{\pi N}=0.4$.}
\end{table}
\vspace{0.4cm}
\begin{table}[hm]
\begin{tabular}{|l|ccc|}
\hline
&&&\\
$~~Q^2$&$A_{1/2}$&$A_{3/2}$&$S_{1/2}$\\
(GeV$^2$)&&&\\
&&&\\
\hline
   1.8&$  -37.5\pm 0.8\pm 1.1$
&$  25.5\pm 0.8\pm 1.8$&$ -8.3\pm 0.9\pm 1.6$\\
   2.2&$  -30.2\pm 0.7\pm 1.7$
&$  22.3\pm 0.7\pm 0.8$&$ -5.7\pm 0.8\pm 1.3$\\
   2.6&$  -25.8\pm 1.2\pm 1.4$
&$  17.8\pm 1.2\pm 1.3$&$ -2.1\pm 1.1\pm 1.1$\\
   3.15&$ -21.3\pm 0.8\pm 2.7$
&$  14.6\pm 0.8\pm 1.8$&$ -0.2\pm 0.7\pm 1.9$\\
   4.0&$ -14.1\pm 0.9\pm 2.7$
&$   8.7\pm 1.1\pm 2.5$&$  1.8\pm 1.2\pm 1.8$\\
\hline
\end{tabular}
\caption{\label{f15_av}
The average values of the $\gamma^* p \rightarrow~N(1680){5\over 2}^+$
helicity amplitudes found using UIM and DR (in units of $10^{-3}$GeV$^{-1/2}$).
The first and second uncertainties are, respectively, the statistical uncertainty from the fit 
and the model uncertainty discussed in the text. The amplitudes are extracted using 
the following mass, width, and $\pi N$ branching
ratio of the resonance: 
$M=1.685~$GeV, $\Gamma=0.13~$GeV, and $\beta_{\pi N}=0.65$.
}
\end{table}
\begin{table}[hm]
\begin{tabular}{|l|cc|}
\hline
&&\\
$~~Q^2$&$A_{1/2}$&$S_{1/2}$\\
(GeV$^2$)&&\\
&&\\
\hline
   1.8&$  19.4\pm 2.4\pm 4.0$
&$ -6.3\pm 2.9\pm 1.1$\\
   2.2&$  9.7\pm 2.2\pm 2.8$
&$ -5.2\pm 2.7\pm 1.1$\\
   2.6&$  -1.2\pm 2.9\pm 2.5$
&$ -6.0\pm 2.6\pm 1.3$\\
   3.15&$ 2.2\pm 2.2\pm 2.6$
&$ -5.6\pm 2.9\pm 1.2$\\
   4.0&$ 2.7\pm 2.3\pm 2.7$
&$  -4.1\pm 3.1\pm 1.4$\\
\hline
\end{tabular}
\caption{\label{p11_av}
The average values of the
$\gamma^* p \rightarrow~N(1710){1\over 2}^+$
helicity amplitudes
found using UIM and DR (in units of $10^{-3}$GeV$^{-1/2}$).
The first and second uncertainties are, respectively, the statistical uncertainty from the fit 
and the model uncertainty discussed in the text.
The amplitudes are extracted using the following mass, width, and $\pi N$ branching
ratio of the resonance: 
$M=1.71~$GeV, $\Gamma=0.1~$GeV, and $\beta_{\pi N}=0.15$.
}
\end{table}

In the global analysis, we have taken into account
all  3- and 4-star resonances
from the first, second, and third resonance
regions.
From the resonances of the fourth resonance region,
we have included the $\Delta(1905){5\over 2}^+$ and the 
$\Delta(1950){7\over 2}^+$. For the masses, widths, and  $\pi N$ branching
ratios  of the resonances, we used
the mean values of the data from the
Review of Particle Physics~\cite{Agashe:2014kda} (see also Table V in Ref. \cite{Azn2009}).
The results on the resonances of
the first and second resonance regions, including their model uncertainties
are based on the data \cite{Park:2007tn}.
They have been found and presented
in Ref. \cite{Azn2009}.
The analysis of the combined sets of data
allowed us to get reliable results
for the electroexcitation amplitudes of the 
following states from the third resonance region:
$N(1675){5\over 2}^-$,
$N(1680){5\over 2}^+$, and $N(1710){1\over 2}^+$.
The isotopic pairs of the resonances from this region:
$\Delta(1600){3\over 2}^+$ and $N(1720){3\over 2}^+$,
$\Delta(1620){1\over 2}^-$ and $N(1650){1\over 2}^-$,
$\Delta(1700){1\over 2}^-$ and $N(1700){1\over 2}^-$,
could not be separated from each other from the data on the $N\pi$ production
in a single channel. For their investigation, data in at least two
channels,
$\gamma^* p\rightarrow n \pi^+$ and $\gamma^* p\rightarrow p \pi^0$, are necessary.
Concerning  
resonances of the fourth resonance region, the present data did not allow us 
to extract reliably their electroexcitation amplitudes. As these are mostly isospin $3 \over 2$
states, for their determination it is essential to include the $p\pi^0$ channel in the analysis.

\subsection{Discussion of global fits} 

\begin{figure*}[t]
\begin{center}
\includegraphics[width=17.cm,height=6.cm]{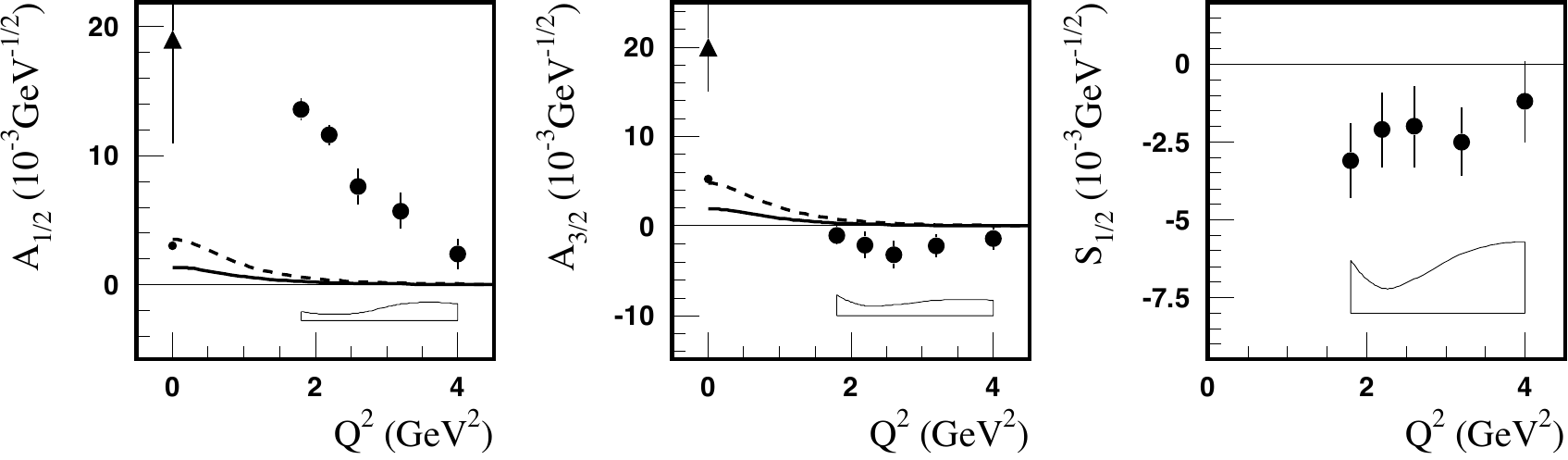}

\caption{\small
Helicity amplitudes
for the  
$\gamma^* p \rightarrow~N(1675){5\over 2}^-$
transition. 
The full circles are
the results from Table \ref{d15_av} obtained
in this work. The bands show the model uncertainties.
The dots at $Q^2=0$ are the predictions of the light-front
relativistic quark model from Ref. \cite{Aznauryan_quark}.
The triangles at $Q^2 = 0$ are the RPP 2014
estimates~\cite{Agashe:2014kda}. 
The dashed and solid curves correspond
to the quark model predictions of Refs. \cite{Merten} and \cite{Giannini}, 
respectively.
\label{d15}}
\vspace{1.cm}
\includegraphics[width=17.cm,height=6.cm]{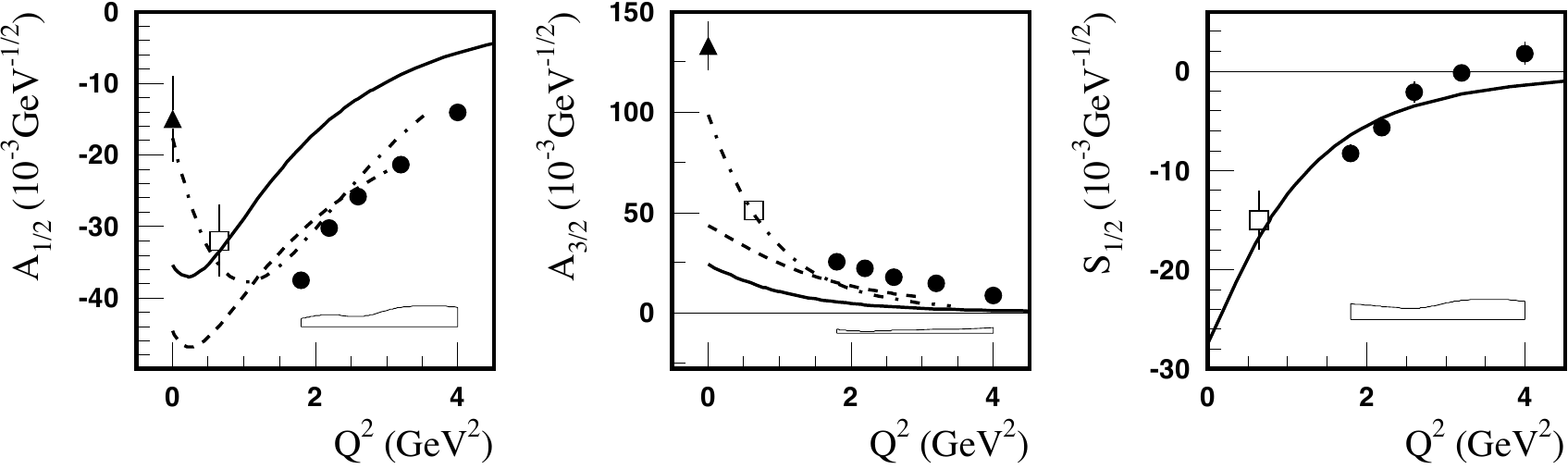}
\caption{\small
Helicity amplitudes
for the  
$\gamma^* p \rightarrow~N(1680){5\over 2}^+$
transition. 
The full circles are
the results from Table \ref{f15_av} obtained
in this work.
The open boxes are  the results of the combined analysis
of CLAS single $\pi$ and 2$\pi$ electroproduction data
\cite{Azn065}.
The full triangles at $Q^2=0$ are the RPP 2014 estimates~\cite{Agashe:2014kda}.
The curves correspond to quark model predictions:  dashed - Ref. \cite{Merten}, 
solid - Ref. \cite{Giannini}, dashed-dotted - Ref. \cite{Close}. 
\label{f15}}
\end{center}
\end{figure*}

\begin{figure*}
\begin{center}
\includegraphics[angle=0,width=0.47\textwidth]{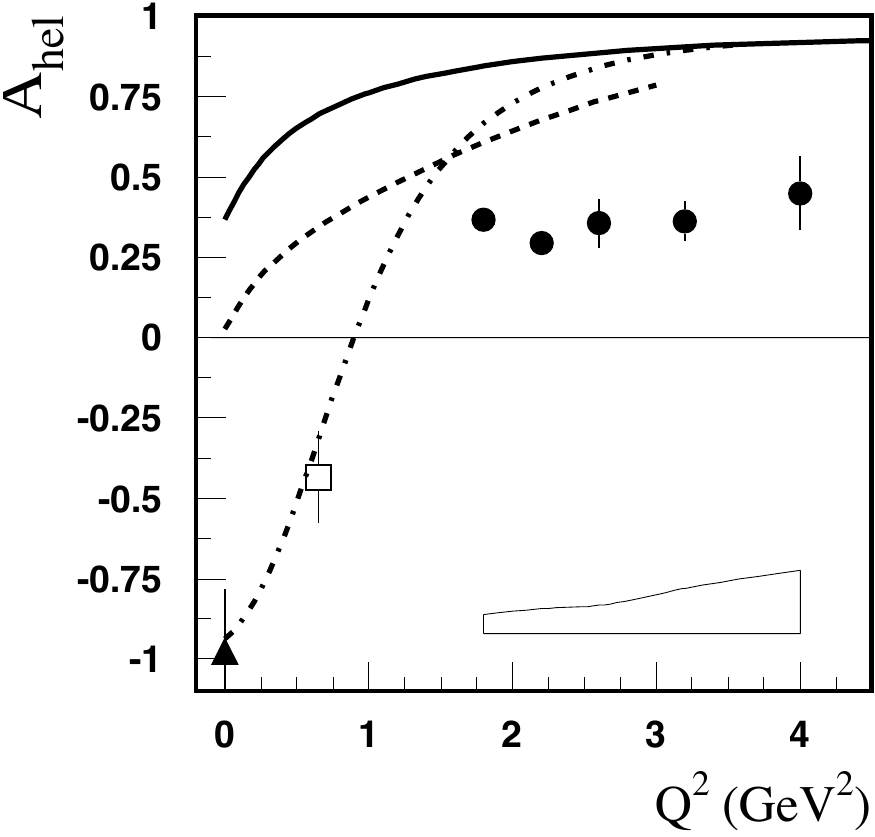}
\caption{\small
Helicity asymmetry 
 $A_{hel}=(A_{1/2}^2-A_{3/2}^2)/(A_{1/2}^2+A_{3/2}^2)$
for the  $\gamma^* p \rightarrow~N(1680){5 \over 2}^+$
transition. The legend is as for Fig. \ref{f15}.
\label{f15_asym}}

\vspace{1.0cm}
\includegraphics[angle=0,width=0.8\textwidth]{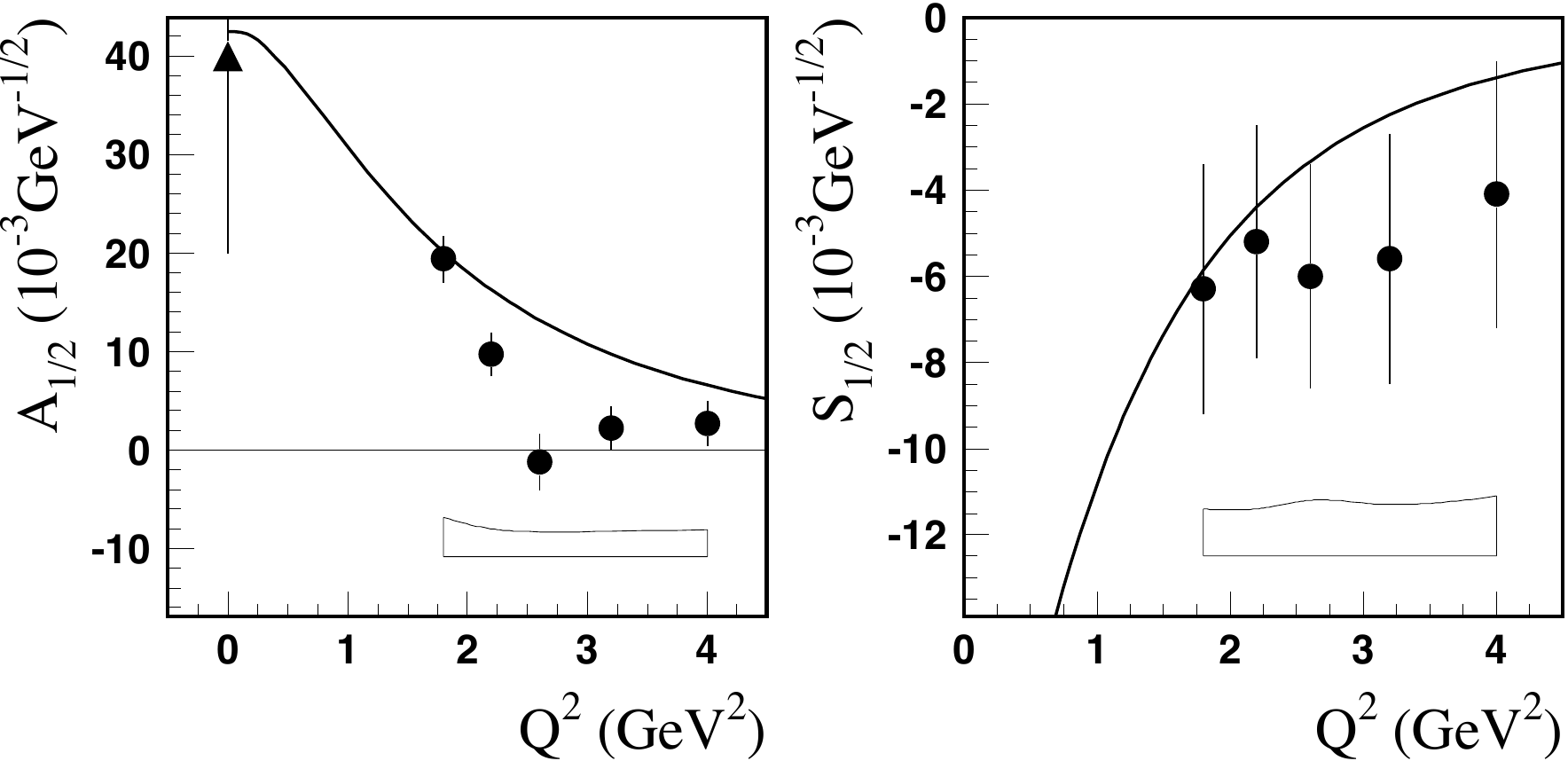}
\caption{\small 
Helicity amplitudes
for the  
$\gamma^* p \rightarrow~N(1710)1/2^+$
transition. 
The full circles are
the results from Table \ref{p11_av} obtained
in this work. 
The solid curves correspond
to quark model predictions of Ref. \cite{Giannini}. 
The legend is as for Fig. \ref{d15}.
\label{p11}}
\end{center}
\end{figure*}

The results 
for the electroexcitation amplitudes of the resonances 
$N(1675){5\over 2}^-$,  $N(1680){5\over 2}^+$, and $N(1710){1\over 2}^+$
are presented in Tables \ref{d15_av}, \ref{f15_av},  and \ref{p11_av},
and Figs. \ref{d15}, \ref{f15}, and  \ref{p11}. 
The presented amplitudes 
are the averaged values of the results obtained using UIM and DR.
The uncertainty that originates from the averaging is considered
as one of the model uncertainties. Following the analysis made
in Ref. \cite{Azn2009}, we consider also two other kinds
of model uncertainties. The first one arises from
the uncertainties of the widths and masses of the resonances.
It is caused mainly by the poor
knowledge of the width of the $N(1710){1\over 2}^+$.
The second one is related to the uncertainties of the background of the UIM
and the Born term in DR. The pion and nucleon electromagnetic form factors
that enter these quantities are known quite well
from experimental data \cite{Melnitchouk,Lachniet,Horn,Tadevos,Riordan},
and the second uncertainty is caused mainly by the poor knowledge of
the $\rho \rightarrow \pi \gamma$ form factor.
According to the QCD sum rule \cite{Eletski}
and the quark model \cite{AznOgan} predictions, the $Q^2$ dependence
of this form factor is
close to the dipole form
$G_D(Q^2) =1/(1+\frac{Q^2}{0.71\rm GeV^2})^2$.
We used this form in our analysis and have introduced in our final results a systematic
uncertainty that accounts for a $20\%$ deviation
from $0.71~$GeV$^2$. All these uncertainties are added in quadrature and presented 
as the model uncertainties of the amplitudes.

\subsection{The $N(1675){5\over 2}^-$ resonance}

The single quark transition model (SQTM), based on the approximation
that only a single quark is involved in the resonance transition,
predicts the suppression of both transverse amplitudes 
$A_{1/2}$ and $A_{3/2}$ for $\gamma^* p \rightarrow~N(1675){5\over 2}^-$
\cite{Hey_Weyers,Babcock_Rosner,Cottingham,SQTM}. 
This suppression is known as the Moorhouse selection rule \cite{Moorhouse}. 
The suppression of the transverse amplitudes $A_{1/2}$ and $A_{3/2}$
for the $N(1675){5\over 2}^-$, predicted by
the SQTM, is confirmed by the results obtained in dynamical quark
models: by the light-front relativistic quark model \cite{Aznauryan_quark} at $Q^2=0$
and in the quark models  \cite{Merten,Giannini} at all $Q^2$ under consideration.
As it can be seen from Fig. \ref{d15}, the suppression of the 
amplitude $A_{1/2}$, as predicted
by the quark models, strongly disagrees with the results 
extracted from the experimental data. We note that these results are independent of
what model was used in the fit: UIM or DR. For the $A_{3/2}$ amplitude we 
observe values slightly negative and consistent with zero within the overall 
uncertainties (statistics + systematics + model), which, if we take the value at 
the photon point ($Q^2=0$) as a reference, shows a much more rapid drop of 
its strength with $Q^2$ compared to $A_{1/2}$.  

Therefore, we can conclude that the transverse amplitudes 
for the transition $\gamma^* p \rightarrow~N(1675){5\over 2}^-$
are determined almost entirely due to non single-quark contributions. 
It should be noted that, in contrast, significant strength through quark 
transition is expected for both transverse amplitudes in the excitation 
of this state from the neutron~\cite{SQTM}.

\subsection{The $N(1680){5\over 2}^+$ resonance}

The amplitudes for the  $\gamma^* p \rightarrow~N(1680){5\over 2}^+$
transition extracted from the experimental data are shown in Fig. \ref{f15}
along with the predictions of quark models:
the relativistic model of Ref. \cite{Merten} and the nonrelativistic models 
\cite{Close,Giannini}. All models underestimate the value of the
amplitude $A_{3/2}$. Also all models predict significant dominance
of the $A_{1/2}$ amplitude over $A_{3/2}$ with increasing $Q^2$, which
is not seen in the amplitudes extracted from the data.
This can be more clearly seen in Fig. \ref{f15_asym} in terms of the helicity
asymmetry, which shows only a very slow rise at $Q^2 > 2$~GeV$^2$. 
A possible explanation of these discrepancies is a large
meson-cloud contribution to the amplitude $A_{3/2}$, which according 
to investigation within a coupled-channel approach~\cite{Sato2008}, can be 
quite significant even above $2~$GeV$^2$. 

\subsection{The $N(1710){1\over 2}^+$ resonance}

This state has a 3* rating in the RPP~\cite{Agashe:2014kda}, and additional confirmation 
from channels other than elastic scattering $\pi N \to \pi N$ is desirable to strengthen 
its status. The current analysis shows the need to include the state into the fit.  For the
two lower $Q^2$ points, finite values of $A_{1/2}$ are extracted, while at the higher $Q^2$ 
the values for $A_{1/2}$ are smaller than the experimental and model 
uncertainties. The $S_{1/2}$ amplitude, although small in magnitude, is negative but 
with finite values that are close to the predictions of a recent quark model calculation
\cite{Giannini}, which also is close to the extracted transverse amplitude $A_{1/2}$ as 
shown in Fig. \ref{p11}.

\section{Conclusions}
For the first time we have measured differential cross sections for the exclusive electroproduction process $ep \to e^\prime \pi^+ n$ in the 
range of the invariant mass of the pion-nucleon system $1.6 \leq W \leq 2.0$~GeV, at photon virtuality 
$1.8 \leq Q^2 < 4.5$~GeV$^2$, and with nearly full coverage in the
azimuthal and polar angles of the $n\pi^+$ center-of-mass system. A total of approximately 37,000 differential 
cross section data points were obtained. This data set, together with the earlier published data set of similar size covering the 
lower mass region $W = 1.1 - 1.69$~GeV, provides complete coverage of the nucleon resonance region up to 
$W = 2$~GeV and $Q^2 < 4.5$~GeV$^2$, that can be used as input for multi-channel partial wave analyses to determine 
the $Q^2$ dependence of electroexcitation of $N^*$ and $\Delta^*$ states with masses up to $1.9-2.0$~GeV. Data for the equivalent 
neutral pion final state $p\pi^0$ that are needed for the separation of isospin $1\over 2$ and isospin $3\over 2$ 
states will be published in the future.    

We employed a single-channel energy-dependent resonance analysis framework in a global fit of 
the 37,000  differential cross section points to extract the helicity amplitudes 
$A_{1/2}$,~$A_{3/2}$,~$S_{1/2}$ and their $Q^2$ dependence for some of the 
well-known isospin $1\over 2$ $N^*$ states. As is true for this type of analysis, our global data fit has 
some model-sensitivity. Much of this sensitivity is due to the uncertainty 
in the non-resonant background amplitudes. In order to have a quantitative measure of the sensitivity to the specific 
modeling of the background amplitudes in the fit we employed  
two independent approaches that describe the background amplitudes in very different ways. These are the unitary 
isobar model and the fixed-t dispersion relation approach. The results are quite consistent and show only 
relatively minor differences in the extracted  helicity amplitudes for the states that are most sensitive to the measured channel and that are relatively 
isolated and have no isotopic partners with similar masses, i.e. $N(1675){5\over 2}^-$, $N(1680){5\over 2}^+$, 
and $N(1710){1\over 2}^+$. 
The latter is the least well determined state as its coupling to $N\pi$ is relatively weak and not well determined. 
For the other two the coupling to $N\pi$ is well measured, and the resonance amplitudes, masses, and hadronic decays 
widths are well determined from elastic $\pi N \to \pi N$ scattering.   

Our data cover the mass range up to 2 GeV, and are thus sensitive to many $N^*$ and $\Delta^*$ 
states. All of these states were used in the global analysis. However, 
the single channel analysis does not allow the separation of the different isospin contributions. We have therefore 
limited our analysis to the determination of those resonances that are most sensitively probed in the $ep \to e^\prime \pi^+n$ 
channel, i.e. $N^*$ states, and do not overlap with $\Delta^*$ states of the same spin and parity. We also restricted 
the analysis to masses below $W=1.8$~GeV. This leaves
the three states for which we show the resulting electrocoupling amplitudes, $N(1675){5\over 2}^-$, $N(1680){5\over 2}^+$,
and $N(1710){1\over 2}^+$. 

The most intriguing result of this analysis is the strong deviation of the $A_{1/2}$ amplitude for the transition to the $N(1675){5\over 2}^-$ from the CQM predictions at all measured $Q^2$. Dynamical quark model predict more than an order of magnitude smaller values than what was extracted from the data. 
To our knowledge this is to 
date the strongest and most direct evidence for dominant non-quark contribution to the 
electroexcitation of a nucleon resonance on the proton. The relative strength of quark contributions and meson-baryon contributions will become much clearer when data on neutrons become available. The analysis of such measurements is underway with data taken on a liquid-deuterium target with CLAS. 

The helicity amplitudes for the $N(1680){5\over 2}^+$ show a transition from $A_{3/2}$ dominance at the
real photon point to $A_{1/2}$ dominance at high $Q^2$. This is a longstanding prediction by the CQM. 
However, the transition is much less rapid than what is predicted, and does not quantitatively agree with 
the constituent quark models, which indicates that for some states non-quark contributions may be relevant 
even at relatively large $Q^2$. It will be very interesting to study the transition amplitudes to even higher 
values of $Q^2$ to see if this trend continues.      

The data set presented in this work has great potential to reveal the internal structure of states 
for which the transition amplitudes could not be quantified using a single channel 
analysis approach. In the near future data will be available from the $e p \to e^\prime p^\prime \pi^0$
 channel, including a variety of single and double polarization asymmetries with polarized beam and targets. These data have high sensitivity to relative phases between different partial waves. Their inclusion into a two-channel analysis will allow for an extraction of the $\Delta^*$ states as well as other $N^*$ states. These studies should also be further extended to higher $Q^2$ where no data 
exist at all, as well as to $Q^2 < 2$~GeV$^2$, where only limited data exist. 
This will allow for the determination of the transition charge and 
current densities of individual states through a Fourier transformation of the transverse amplitudes in the 
light cone frame. Such data can reveal novel information of the internal structure of the excited states in transverse impact parameter space~~\cite{Carlson:2007xd,Tiator:2008kd}.

\section{Acknowledgments}

We are grateful for the efforts of the staff  of the Accelerator and Physics Divisions at Jefferson Lab that
made this experiment possible.  
This material is based upon work supported by the U.S. Department of Energy, 
Office of Science, Office of Nuclear Physics under contract DE-AC05-06OR23177. 
This work was also supported by the 
National Science Foundation, the State Committee of Science of Republic 
of Armenia Grant 13-1C023, the Chilean Comisi\'on Nacional de Investigaci\'on Cientifica 
y Tecnol\'ogica (CONICYT), the Italian Istituto Nazionale di Fisica Nucleare, the 
French Centre National de la Recherche Scientique, the French Commissariat a 
l'Energie Atomique,  the Scottish Universities Physics Alliance (SUPA), the 
United Kingdom's Science and Technology Facilities Council, and the National Research 
Foundation of Korea.


\begin{thebibliography}{999}

\bibitem{isgur_2000} N. Isgur, nucl-th/0007008 (2000), 
in: Excited Nucleons and Hadronic Structure, World Scientific, 
eds: V. Burkert,~L. Elouadrhiri,~J. Kelly, R. Minehart. 

\bibitem{Edwards:2011jj}  R.~G.~Edwards, J.~J.~Dudek, D.~G.~Richards and S.~J.~Wallace,
  Phys.\ Rev.\ D {\bf 84}, 074508 (2011).
   
\bibitem{Joo2002} K. Joo {\em et al.}, 
Phys. Rev. Lett. {\bf 88}, 122001 (2002).

\bibitem{Joo2003} K. Joo {\em et al.}, 
Phys. Rev. C {bf 68}, 032201 (2003). 

\bibitem{Sparveris2003} N. F. Sparveris {\em et al.}, 
Phys. Rev. C {\bf 67}, 058201 (2003).

\bibitem{Biselli2003}A. Biselli {\em et al.}, 
Phys. Rev. C{\bf  68}, 035201 (2003).

\bibitem{Kelly2005} J. Kelly {\em et al.}, 
Phys. Rev. Lett. {\bf 95}, 102001 (2005).  

\bibitem{Stave06} S. Stave {\em et al.} 
Eur. Phys. J. A {\bf 30}, 471-476 (2006).

\bibitem{Ungaro06} M. Ungaro {\em et al.}, 
Phys. Rev. Lett. {\bf 97}, 112003 (2006).

\bibitem{Sparveris06} N.F. Sparveris {\em et al.}, 
[arXiv:nucl-ex/0611033].

\bibitem{Biselli2008}A. Biselli {\em et al.}, 
Phys. Rev. C {\bf 78}, 045204 (2008).

\bibitem{Joo2004} K. Joo {\em et al.}, 
Phys. Rec. C{\bf  70}, 042201 (2004).  

\bibitem{Egiyan:2006ks} 
  H.~Egiyan {\it et al.},
  Phys. Rev. C {\bf 73}, 025204 (2006).

\bibitem{Park:2007tn} K. Park {\it et al.}, 
Phys. Rev. C $\bf{77}$, 015208 (2008). 

\bibitem{Armstrong1999} C. Armstrong {\em et al.},
 Phys. Rev. D {\bf 60}, 052004 (1999).

\bibitem{Thompson2001} R. Thompson {\em et al.}, 
Phys. Rev. Lett. {\bf 86}, 1702 (2001).

\bibitem{Denizli:2007tq} 
  H.~Denizli {\it et al.},
  Phys. Rev. C {\bf 76}, 015204 (2007).

\bibitem{Aznauryan:2008pe} 
  I.~G.~Aznauryan {\it et al.},
  Phys. Rev. C {\bf 78}, 045209 (2008).
  
\bibitem{Fedotov:2008aa} 
  G.~V.~Fedotov {\it et al.},
   Phys. Rev. C {\bf 79}, 015204 (2009).

\bibitem{Mokeev:2008iw} 
  V.~I.~Mokeev, V.~D.~Burkert, T.~-S.~H.~Lee, L.~Elouadrhiri, G.~V.~Fedotov and B.~S.~Ishkhanov,
  Phys.\ Rev.\ C {\bf 80}, 045212 (2009).

\bibitem{Mokeev:2012vsa} 
  V.~I.~Mokeev {\it et al.},
  Phys.\ Rev.\ C {\bf 86}, 035203 (2012).
  
\bibitem{Aznauryan2012} I. G. Aznauryan and V. D. Burkert,
  Prog. Part. Nucl. Phys. {\bf 67}, 1 (2012).
  
\bibitem{Aznauryan2013}
 I. G. Aznauryan {\it et al.}, Int. J. Mod. Phys. E {\bf 22} (2013) 1330015.  

\bibitem{burkert_lee_2004} V.D. Burkert and T.-S. H. Lee, 
Int. J. Mod. Phys. E {\bf 13}, 1035 (2004).

\bibitem{Aznauryan:2012ec} 
  I.~G.~Aznauryan and V.~D.~Burkert, 
  Phys.\ Rev.\ C {\bf 85}, 055202 (2012).
  
\bibitem{Ramalho:2013mxa} 
 G.~Ramalho and M.~T.~Pena,
 Phys.\ Rev.\ D {\bf 89}, 094016 (2014).

\bibitem{Thomas} D. H. Lu, A. W. Thomas, and A. G. Williams,
Phys. Rev. C {\bf 55} 3108 (1997).

\bibitem{Faessler} A. Faessler, T. Gutsche, B.R. Holstein et al.,
Phys. Rev. D {\bf 74} 074010 (2006).

\bibitem{Kamalov1999} S.S. Kamalov and S.N. Yang,
Phys. Rev. Lett. {\bf 83} 4494 (1999).

\bibitem{Kamalov2001} S.S. Kamalov {\it et al.},
Phys. Rev. C {\bf 64} (R) 032201 (2001).
  
\bibitem{Sato2001} T. Sato and T.-S. H. Lee,
Phys. Rev. C {\bf 63} 055201 (2001).

\bibitem{Sato2007}  A. Matsuyama, T. Sato, and T.-S.H. Lee,
Phys. Rep. {\bf 439} 193 (2007).

\bibitem{Sato2008} B. Juli\'a-D\'iaz, T.-S.H. Lee, A. Matsuyama, and T. Sato,
Phys. Rev. C {\bf 77} 045205 (2008).

\bibitem{Ramalho:2011ae} 
  G.~Ramalho and M.~T.~Pena,
  Phys.\ Rev.\ D {\bf 84}, 033007 (2011).

\bibitem{Ramalho:2010js} 
  G.~Ramalho and K.~Tsushima,
  Phys.\ Rev.\ D {\bf 81}, 074020 (2010).
  
\bibitem{Agashe:2014kda} 
  K.~A.~Olive {\it et al.}  [Particle Data Group Collaboration],
  Chin.\ Phys.\ C {\bf 38}, 090001 (2014). 

\bibitem{clas} B.A. Mecking {\em et al.}, 
Nucl. Instr. and Meth. A {\bf 503}, 513 (2003).

\bibitem{geant3} R. Brun and F. Rademaker, 
Nucl. Instr. Meth. A {\bf 389}, 81 (1997).

\bibitem{MAID} D. Drechsel, O. Hanstein, S.S. Kamalov 
and L. Tiator, Nucl. Phys. A {\bf 646}, 145 (1999)
http://www.khp.uni-mainz.de/MAID/. 

\bibitem{maid2007} D. Drechsel, O. Hanstein, S. Kamalov, and L. Tiator, 
Eur. Phy. J. A {\bf 34}, 69 (2007).

\bibitem{Afanasev:2002ee}
  A.~Afanasev, I.~Akushevich, V.~Burkert and K.~Joo,
  Phys.\ Rev.\ D {\bf 66}, 074004 (2002).

\bibitem{clas_db} CLAS Physics Database, 
 http://clasweb.jlab.org/physicsdb.
 
\bibitem{klam2013} D.~S.~Carman, K.~Park {\em et al.},
Phys.~Rev.~C\textbf{87}, 025204 (2013).

\bibitem{Azn2003} I. G. Aznauryan,
Phys. Rev. C $\bf{67}$, 015209 (2003).

\bibitem{Azn2009} I. G. Aznauryan {\it et al.},
Phys. Rev. C $\bf{80}$, 055203 (2009).

\bibitem{Azn2005} I. G. Aznauryan, V. D. Burkert, H. Egiyan,
{\it et al.}, Phys. Rev. C $\bf{71}$, 015201 (2005).

\bibitem{Azn065} I. G. Aznauryan, V. D. Burkert,
{\it et al.}, Phys. Rev. C $\bf{72}$, 045201 (2005).

\bibitem{Crawford}  R. L. Crawford,
Nucl. Phys. B $\bf{97}$, 125 (1975).

\bibitem{Melnitchouk} J. Arrington, W. Melnitchouk, J. A. Tjon,
Phys. Rev. C {\bf 76}, 035205 (2007).

\bibitem{Lachniet} J. Lachniet {\em  et al.}, 
Phys. Rev. Lett. {\bf 102}, 192001 (2009).

\bibitem{Horn} T. Horn  {\it et al.}, Phys. Rev. Lett. {\bf 97}, 192001
(2006).

\bibitem{Tadevos} V. Tadevosyan {\it et al.}, Phys. Rev. C  {\bf 75},
055205 (2007).

\bibitem{Riordan} S.Riordan {\em et al.}, Phys. Rev. Lett.
{\bf 105}, 262302 (2010).

\bibitem{Eletski} V. Eletski and Ya. Kogan, Yad. Fiz. {\bf 39}, 138
(1984).

\bibitem{AznOgan} I. Aznauryan and K. Oganessyan, Phys. Lett. B
{\bf 249}, 309 (1990).

\bibitem{Hey_Weyers} A.J.G. Hey and J. Weyers,
Phys. Lett. B {\bf 48}, 69 (1974).

\bibitem{Babcock_Rosner} J. Babcock and J.L. Rosner,
Ann. Phys. (N.Y.)  {\bf 96}, 191  (1976) .

\bibitem{Cottingham} W.N. Cottingham and I.H. Dunbar,
Z. Phys.  C {\bf 2}, 41 (1979) .

\bibitem{SQTM} V.D. Burkert {\it et al.},
Phys. Rev. C {\bf 67}, 035204 (2003) .

\bibitem{Moorhouse} R. G. Moorhouse,
Phys. Rev. Lett. {\bf 16}, 771 (1966).

\bibitem{Aznauryan_quark} I. G. Aznauryan and A. S. Bagdasaryan,
Yad. Fiz. {\bf 41}, 249 (1985); translation in Sov. J. Nucl. Phys. $\bf{41}$, 158 (1985).

\bibitem{Merten} D. Merten, U. L\"oring, B. Metsch, and H. Petry,
Eur. Phys. J. A {\bf 18}, 193 (2003).

\bibitem{Giannini} E. Santopinto and M. M. Giannini,
Phys. Rev. C {\bf 86}, 065202 (2012).

\bibitem{Close} Z. Lee and F. E. Close, 
Phys. Rev. D {\bf 42}, 2207 (1990).

\bibitem{Carlson:2007xd}  C.~E.~Carlson and M.~Vanderhaeghen,
  Phys.\ Rev.\ Lett.\  {\bf 100}, 032004 (2008).

\bibitem{Tiator:2008kd}  L.~Tiator and M.~Vanderhaeghen,
  Phys.\ Lett.\ B {\bf 672}, 344 (2009).

\end{thebibliography}
\end{document}